# Sequence of bifurcations of natural convection of air in a laterally heated cube with perfectly insulated horizontal and spanwise boundaries

Alexander Gelfgat

School of Mechanical Engineering, Faculty of Engineering, Tel-Aviv University, Ramat Aviv, Tel-Aviv, Israel, 6997801, gelfgat@tau.ac.il

**Abstract**

A sequence of three steady – oscillatory transitions of buoyancy convection of air in a laterally heated cube with perfectly thermally insulated horizontal and spanwise boundaries is studied. The problem is treated by Newton and Arnoldi methods based on Krylov subspace iteration. The finite volume grid is gradually refined from $100^3$ to $256^3$ finite volumes. It is shown that the primary instability is characterized by two competing eigenmodes, whose temporal development results in two different oscillatory states that differ by their symmetries. Bifurcations due to both modes are subcritical. These modes develop into different oscillatory and then stochastic flow states, which, at larger Grashof number, stabilize and arrive to single stable steady flow. With further increase of the Grashof number this flow loses it stability again. It is argued that in all the three transitions, the instabilities onsets, as well as reinstatement of stability, take place owing to an interaction between a destabilizing centrifugal mechanism and stabilizing effect of thermal stratification.

## 1. Introduction

A primary goal of this study is to examine a chain of three steady – oscillatory bifurcations that take place in a three-dimensional benchmark configuration of natural convection of air in a laterally heated cube with perfectly insulated horizontal and spanwise boundaries. The two-dimensional analog of this problem, i.e., convection in a square laterally heated cavity with perfectly thermally insulated (adiabatic) horizontal boundaries, is one of the earliest CFD benchmarks [1] proposed for validation of steady flow calculations. It was then extended for comparison of calculated critical parameters of the primary bifurcation corresponding to steady to oscillatory transition. The details and additional references can be found in [2]. Later, the two-dimensional problem was extended to a three-dimensional one in a cube, assuming the spanwise boundaries to be also perfectly thermally insulated. The results for 3D steady flows have been published and cross-verified [3-8], so that the 3D steady flows benchmark is also well established. However, numerical investigation of the instabilities of 3D steady flows is significantly more challenging and had been addressed only by straight-forward time integration of the 3D time-dependent governing equations [9-14].

Recently, a comprehensive linear stability analysis was applied to the primary steady - oscillatory transition of air convection in a laterally heated cube in two simpler configurations, for a cube with perfectly thermally conducting horizontal and spanwise boundaries [15], and with perfectly insulated horizontal and perfectly conducting spanwise boundaries [16]. The second case appeared to be noticeably more complicated because the critical Grashof number becomes almost two orders of magnitude larger, and the steady – oscillatory transition is preceded by a symmetry breaking steady bifurcation. A replacement of the perfectly conducting spanwise boundaries by the perfectly insulated ones leads to a qualitatively different transition, which exhibits a sequence of bifurcations described below. This finding is in line with recent computational and experimental study [17], where the authors concluded that the flow can be stabilized or destabilized by variation of the heat transfer conditions at the horizontal boundaries.

The time-dependent calculations of [14] showed that with the increase of the Grashof number $Gr$, the steady flow becomes oscillatory unstable at $Gr > 4 \cdot 10^7$, then the stability reinstates at $Gr > 7 \cdot 10^7$, and the resulting steady flow becomes oscillatory unstable at $Gr > 2 \cdot 10^8$. Following a series of the cited above previous studies, we consider the fixed value of the Prandtl number, $Pr = 0.71$, characteristic for air. The flows with very large [18] and very small [19,20]

Prandtl numbers were also considered for the current configuration, however dependence of the transition on the Prandtl number is out of scope of the present study.

In this study, we examine the instabilities by applying linear stability analysis, using the numerical approach and the visualization technique of [15,16]. For calculation of steady flows we apply the Newton method, whose corrections are calculated by the biconjugate gradient stabilized (BiCGstab(2)) method combined with generalized minimal residual (GMRES) method. Leading eigenvalues are computed by the Chebyshev preconditioned Arnold iteration [21]. The corresponding Krylov vectors, which are divergence free and satisfy all the boundary conditions are computed by the SIMPLE-like technique proposed in [15,22].

Applying the linear stability analysis to the calculated 3D steady flows, we confirm three transitions predicted by the time-dependent computations [14]. We discuss which physical mechanisms destabilize the flow, then stabilize it, and then destabilize it again. Along with that, we confirm existence of two different most unstable modes of the primary instability, as was predicted by [11]. We show that these modes differ by broken flow symmetries and become unstable at close Grashof numbers. Additional time-dependent calculations revealed that beyond the stability limits, there exist two different oscillatory flow states, which also differ by their symmetries. It was shown also that the steady – oscillatory transitions caused by any of the two modes are subcritical. With further increase of the Grashof number, these modes turn into stochastic regimes with different phase space attractors. Surprisingly, when the Grashof number is increased to approximately $7 \times 10^7$, the two attractor collapse into the same stable focus, thus producing a single stable steady flow state. This flow state remains stable up to the Grashof number $\approx 2.9 \times 10^8$, after which the flow becomes oscillatory and then turbulent.

Analyzing the most unstable disturbances and steady flow patterns near the critical points, we argue that in all three cases the destabilization and stabilization result from an interaction of two main factors: destabilizing centrifugal instability mechanism and stabilizing thermal stratification. The centrifugal instability sets in at the cube corners where the flow direction turns from horizontal to vertical, and vice versa, similarly to what is observed in the 3D lid driven cavity flow (see [22] and references therein). This mechanism can be enhanced by reverse circulations that appear near the top and bottom borders, and increase the curvature of streamlines of the main convective circulation. On the other hand, a strong convective mixing tends to make the isotherms far from the vertical isothermal boundaries almost horizontal, so that colder and heavier fluid is located below the warmer and lighter one. This results in a stable stratification, which tends to suppress all possible instabilities.

## 2. Formulation of the problem

As in our previous studies [14-16], we consider the natural convection in an incompressible fluid in a cubic cavity, whose opposite sidewalls are maintained at constant but different temperatures, $T_{hot}$ and $T_{cold}$. The horizontal and spanwise boundaries are perfectly thermally insulated, as is defined for the AA – AA case in [14]. The Boussinesq approximation is applied. To render equations dimensionless we choose $H^2/\nu$, $\nu/H$, $\rho\nu^2/H^2$ as scales of the length, time $t$, the velocity $\boldsymbol{v} = (u, v, w)$ and the pressure $p$, respectively, where $\nu$ is the fluid kinematic viscosity and $\rho$ is the density. The temperature is rescaled to a dimensionless function by $T \rightarrow (T - T_{cold})/(T_{hot} - T_{cold})$. Additionally, the dimensionless time, velocity and pressure are scaled, respectively, by $Gr^{-1/2}$, $Gr^{1/2}$, and $Gr$, where $Gr = g\beta(T_{hot} - T_{cold})H^3/\nu^2$ is the Grashof number, $g$ is the acceleration due to gravity, and $\beta$ is the thermal expansion coefficient. The resulting system of energy, momentum, and continuity equations is defined in a cube $0 \leq x, y, z \leq 1$ as:

$$\frac{\partial T}{\partial t} + (\boldsymbol{v} \cdot \nabla)T = \frac{1}{PrGr^{1/2}} \Delta T \tag{1}$$

$$\frac{\partial \boldsymbol{v}}{\partial t} + (\boldsymbol{v} \cdot \nabla)\boldsymbol{v} = -\nabla p + \frac{1}{Gr^{1/2}} \Delta \boldsymbol{v} + T\boldsymbol{e}_z \tag{2}$$

$$\nabla \cdot \boldsymbol{v} = 0 . \tag{3}$$

where $Pr = \nu/\alpha$ is the Prandtl number, and $\alpha$ is the thermal diffusivity. All the boundaries are assumed to be no-slip. Two vertical boundaries at $x = 0,1$ are kept isothermal, so that

$$T(x = 0, y, z) = 1, \quad T(x = 1, y, z) = 0. \tag{4}$$

The absence of the heat flux at the horizontal and spanwise boundaries yields:

$$\left(\frac{\partial T}{\partial y}\right)_{y=0} = \left(\frac{\partial T}{\partial y}\right)_{y=1} = 0 \tag{5}$$

$$\left(\frac{\partial T}{\partial z}\right)_{z=0} = \left(\frac{\partial T}{\partial z}\right)_{z=1} = 0 \tag{6}$$

For discussion purposes, the areas adjacent to the cube edges (0,*y*,0), (0,*y*,1), (1,*y*,0) and (1,*y*,1) are called below the lower left, upper left, lower right, and upper right corners, respectively.

The problem is additionally characterized by three symmetries [11]: (i) reflection symmetry with respect to the midplane $y = 0.5$, $\{u, v, w, \theta\}(x, y, z) = \{u, -v, w, \theta\}(x, 1 - y, z)$, (ii) 2D rotational symmetry with respect to rotation through 180° around the line $x = z = 0.5$, $\{u, v, w, \theta\}(x, y, z) = -\{u, -v, w, \theta\}(1 - x, y, 1 - z)$, and (iii) 3D centro-symmetry $\{u, v, w, \theta\}(x, y, z) = -\{u, v, w, \theta\}(1 - x, 1 - y, 1 - z)$, where, $\theta = T - (1 - x)$. Obviously, these symmetries are characteristic of steady state flows at relatively small subcritical Grashof numbers. They can be broken by instability, so that supercritical oscillatory flows can maintain only one type of symmetry or be fully non-symmetric. Following [16], to examine whether the steady state solution is symmetric, we use the temperature field and define

$$R_{ref} = max|\theta(x, y, z) - \theta(x, 1 - y, z)| \tag{7}$$

$$R_{2D} = max|\theta(x, y, z) + \theta(1 - x, y, 1 - z)| \tag{8}$$

$$R_{3D} = max|\theta(x, y, z) + \theta(1 - x, 1 - y, 1 - z)| \tag{9}$$

as measures of the reflection, 2D rotational and 3D centro-symmetries, respectively. For the symmetric solutions, these values are of the order of $10^{-9}$. Deviation of any of these expressions from numerical zero indicates symmetry breaking.

## 3. Numerical and visualization techniques

The numerical approach is the same as in [14-16], where the reader is referred for the details. Here we only mention that this approach allows for calculation of divergence free Krylov vectors satisfying all the boundary conditions, which, in its turn, allows for a direct calculation of the leading eigenvalue [21], without applying the Arnoldi method in shift-and-inverse mode like in [23].

As mentioned in [16], convergence of the critical parameters for a similar 2D case of convection in a laterally heated square cavity with insulated horizontal boundaries was established in [23], the convergence and comparison with independent results for 3D steady states were reported in [8], and the convergence of slightly supercritical oscillatory states was studied in [14].

Three-dimensional velocity fields are visualized by the method proposed in [24,25] and applied to similar problems in [15,16], where all the definitions and computational details are discussed. Here, to clarify the plots below, we only mention that 3D flow is visualized by projecting the velocity field on three sets of coordinate planes, (x,y), (y,z), and (x,z), demanding that the projections are divergence free. Each of the three divergence free projections has only two non-zero components along the coordinates of the projection plane, which allows one to define a stream function for each of them. As a result, the 3D velocity field is visualized as isosurfaces of the three scalar functions $\Psi_x$, $\Psi_y$, and $\Psi_z$, which are non-zero components of the vector potentials of the above projections. The visualization of a divergence-free 3D velocity field consists of three independent frames depicting the vector potentials and the velocity projections. The whole visualization approach can be considered as an extension of a two-dimensional stream function to three directions, in which the velocity projection vectors are tangent to the isosurfaces of the corresponding vector potential.

Before discussing the 3D flow patterns, we recall that in a laterally heated 3D cavity the flow is always three-dimensional [2,19,26] It is characterized by a strong convective circulation in the planes located between the isothermal boundaries, and a noticeable weaker motion in the two other sets of vertical and horizontal planes. The weaker motion in the spanwise direction is driven by the pressure drop between the vertical center plane and the spanwise boundaries. At the center plane the velocities are expected to be maximal, vanishing towards the spanwise boundaries due to the no-slip boundary conditions. Consequently, the pressure grows from the center to the spanwise boundaries, which results in the flow from the center towards these boundaries.

Figures 1–3 illustrate the above visualization approach for steady flows at three values of the Grashof number $Gr = 4.4 \times 10^7$ (Fig. 1), $Gr = 7.2 \times 10^7$ (Fig. 2), and $Gr = 2.9 \times 10^8$ (Fig. 3), which are close to the critical values of the three transitions discussed below. In all the three cases the steady flows preserve all the three symmetries. Similarly to what was discussed in Gelfgat (2017, 2020b), the main convective circulation is created by motion in the $(x, z)$ planes and is represented by the vector potential $\Psi_y$: the hot air ascends and the cold air descends along its isosurfaces. Frames (c) in Figs. 1-3 show that at large Grashof numbers central part of the main circulation splits into two (Figs. 1c and 2c) and then into 3 (Fig. 3c) separate circulations. Motion in the $(y, z)$ planes is represented by the isosurfaces of the potential $\Psi_x$, which show that this motion takes place mainly in the boundary layers adjacent to the isothermal boundaries $x = 0$ and 1. Similarly, the motion in the $(x, y)$ planes is visualized by the isosurfaces of the potential $\Psi_z$. This motion is located mainly near the horizontal boundaries, showing existence of a boundary layer

also there. Arrow plots in Figs. 1-3 are tangential to the isosurfaces and illustrate the direction of the divergence-free velocity projections.

It should be noted additionally that unlikely the AA – CC case [14,16] of perfectly conducting spanwise boundaries, the temperature isosurfaces pattern does not exhibit any boundary layers or other steep dependencies near the perfectly insulated spanwise boundaries. Thus, in the central plane $x = 0.5$ the isotherms are almost horizontal and exhibit a visible waviness only near the horizontal boundaries. We observe also that the intensive convective motion brings the cold liquid below and the warm liquid above, so that the isotherms shown in Figs. 1a-3a exhibit a clear stable stratification, which is expected to affect the flow stability.

## 4. Results

### 4.1. Critical Grashof numbers and oscillation frequencies

Convergence of the critical Grashof numbers and critical frequencies was studied on gradually refined grids varied from 100×100 to 256×256 finite volumes. The grids were stretched near the borders as in [14-16]. The critical values corresponding to the three consequent bifurcations exhibited a linear dependence on the squared average size grid, similarly to what is illustrated in Fig. 4 of [15]. This allowed us to perform the Richardson extrapolation [27] and to find zero grid size limits of the critical Grashof numbers and critical frequencies. All the results are listed in Table 1.

The primary bifurcation takes place at $Gr_{cr} \approx 4.50 \cdot 10^7$. The corresponding most unstable perturbation exhibits broken reflection symmetry and 3D centrosymmetry, however preserves the 2D rotational symmetry. Along with this most unstable perturbation mode, we observed another mode, which becomes unstable at $Gr \approx 4.52 \cdot 10^7$. It preserves 3D centrosymmetry, and breaks the two other symmetries. The pertaining of the 3D centrosymmetry with the two other symmetries broken was already observed and discussed in [14,16]. The critical Grashof numbers of these two modes are so close that to arrive to a correct critical value it was necessary to observe four leading eigenvalues. These two modes were observed before in [11], but other studies reported only one of them.

The computed critical parameters are compared with previously published results in Table 1. Parameters of the second and the third bifurcation can be compared only with the results of [14]

Gelfgat (2017). Taking into account difficulties in estimating of the critical Grashof number from results of a time-dependent calculation, the comparison is quite good. The present critical Grashof numbers corresponding to the two modes of the primary bifurcation coincide within the second decimal place with the independent calculations [11], while only the first decimal place is the same for other results. The scatter is noticeably larger in the reported oscillation frequencies. The discrepancy between the present results and those of time-dependent computations of [11], possibly can be explained by change of the oscillations period in slightly supercritical flows. The oscillation frequencies reported in other studies can result either from interaction of the two unstable modes, or from the subcriticality of the primary bifurcation, discussed below. Note, that the dimensionless frequency $\approx 0.01$, reported in the three studies [9,11,14], can be interpreted as one resulting in finite amplitude oscillations of the second primary mode, whose critical frequency obtained here is $\approx 0.008$.

4.2. Primary steady-oscillatory transition

*4.2.1. Perturbations of temperature: comparison with 2D case and fully non-linear time-dependent computations*

As mentioned above, the primary steady – oscillatory transition takes place at $Gr_{cr} \approx 4.5 \times 10^7$ owing to a 2D rotational symmetry preserving, and broking two other symmetries, perturbation. This is accompanied by another, 2D rotational symmetry and reflection symmetry breaking eigenmode, which becomes unstable at $Gr \approx 4.52 \times 10^7$. Starting from a quite small supercriticality, e.g. $Gr = 4.55 \times 10^7$, there exist two competing eigenmodes, so that one can expect to observe interaction of the two modes, as well as two distinct oscillatory states developing independently owing to either of the modes.

Before addressing the slightly supercritical flow states, we compare the 3D instability of the considered flow with its two-dimensional counterpart (see, e.g., [23]), and with the distribution of oscillations amplitude in a fully developed slightly supercritical 3D oscillatory state [14]. Figure 4a-d shows perturbations of the temperature of both modes as isosurfaces (frames a,c), and isolines in the central plane $y = 0.5$ and two neighbor planes $y = 0.45$ and 0.45 (frames b,d). For the comparison purposes we show also amplitudes of the oscillatory 2D flow (Fig. 4e) and of the most unstable 2D disturbance (Fig. 4f). The 3D patterns should be compared with those plotted in Fig.

12 of [14]. We observe that the amplitudes of the perturbations and the oscillations in all the 2D and 3D cases are located in the same corners $x = z = 0$ and $x = z = 1$, and exhibit quite similar patterns. At the same time, one can observe different structures shape in the 2D and 3D disturbances. Moreover, in the case of similar 2D and 3D instability mechanisms, the similarity is expected to be observed in the center plane $y = 0.5$, like in [15]. In the present case the temperature disturbance is zero at the center plane (Fig. 4b and 4c), and attains the maximal values in the nearby planes. Comparing also isosurfaces of the amplitudes of both modes with the oscillation amplitude of the supercritical oscillatory state (Fig. 12a of [14]), we do observe some similarity. The difference in the patterns can be attributed to the changes due to non-linear development of the oscillatory flow. For example, appearance of a noticeable oscillations in the two opposite corners $x = 0, z = 1$ and $x = 1, z = 0$ can trigger strong non-linear effect. The similarity is seen better in the cross-sectional planes (cf. Figs. 4b and 5d with Fig 12c of [14]). There is also certain similarity between the discussed amplitudes of perturbations and oscillatory flows with the POD empirical eigenfunctions reported in Fig. 3 of [28].

*4.2.2. Two most unstable perturbation modes*

For a better comparison of the two modes and an illustration of their symmetry properties, their real and imaginary parts are plotted in Fig 5. We observe that within a phase shift, the two modes exhibit similar patterns that consist of two or four pairs of minima and maxima located in the corners $x = z = 0$ and $x = z = 1$. It is clearly seen that the first mode preserves the 2D centrosymmetry, while breaks the 3D centrosymmetry. Contrarily, the second mode preserves the 3D centrosymmetry and breaks the 2D rotational symmetry. Both modes are antisymmetric with respect to the center plane $y = 0.5$, meaning that both of them break the reflection symmetry. The time evolution of the two disturbances over the oscillations period is visualized in the animation (Animation1 and Animation2). Note also that the real and imaginary parts of the first mode (Fig. 5a,b) are well compared with those of the dominant DMD (dynamic mode decomposition) mode reported in Fig. 4 of [13].

Both most unstable modes have largest amplitudes of the three velocity components in the lower left and upper right corners $x = z = 0$ and $x = z = 1$, where the temperature perturbation amplitude is also largest (Fig. 5). Similarly to the temperature disturbances, the oscillations of the velocity perturbation in time are located near these corners and do not penetrate further in the bulk of the flow (see Animation1 and Animation2). Since all the perturbations of both modes are located close to the isothermal boundaries $x = 0$ and $x = 1$, the corresponding boundary layers can play an important role in the onset of observed instabilities. A closer look at the steady flow patterns reveals that, indeed, there are extremely thin boundary layers of the vertical velocity and the temperature. The corresponding profiles in the spanwise midplane $y = 0.5$ are shown in Fig. 6. We observe the steepest change of the vertical velocity near the horizontal midplane $z = 0.5$ (Fig. 7a,b), while the steepest change of the temperature takes place near the isothermal boundaries (pink and green lines in Fig. 6c).

The observed boundary layers resemble the problem of natural convection in the boundary layer adjacent to a vertical heated plate [29]. Stability of this flow was examined in several studies [30,31], where only two-dimensional perturbations were taken into account. This can be justified by the Squire transformation, which is applicable to parallel natural convection flows. In the present case, we observe three-dimensional most unstable disturbances (Fig. 5), which makes the above results inapplicable for explanation of the observed instability. It is still possible that at large Grashof numbers the most unstable perturbation of the parallel boundary layer flow will become three-dimensional, similarly to what was observed for non-isothermal mixing layer flows at large Richardson numbers in [32,33]. However, even in this case, one would expect disturbances distributed along the vertical boundaries, and not just localized in the corners.

An additional understanding of mechanism that triggers the instability can be gained from the snapshots of the vector potentials showing the divergence-free projections of the velocity disturbance on the coordinate planes. These are also shown for the first and second modes in Figs. 7 and 8, respectively. The snapshots are plotted for the beginning and for a quarter of an oscillation period. The corresponding patterns for the half and three quarters of the period can be obtained by reversing the colors. The corresponding animations, Animation3 and Animation4, show 40 snapshots over an oscillation period. The projected motions are also depicted by their streamlines. The direction of motion on the streamlines changes to the opposite during the oscillations period.

The perturbation patterns shown in Figs. 5, 7, and 8 allow us to speculate about physics of the two self-sustained oscillatory processes. First, we argue, that owing to the base flow symmetries, the flow areas adjacent to the corners $x = z = 0$ and $x = z = 1$ are equivalent with respect to the flow stability properties there. Therefore, the instability develops in the both corners simultaneously, as is seen from the most unstable perturbation patterns (Fig. 5). Second, we observe (see. e.g., Fig. 5) that both perturbations modes are antisymmetric with respect to the midplane $y = 0.5$, meaning that the reflection symmetry will always be broken by the instability. Third, the temperature disturbances shown in Fig. 5 can appear in phase with respect to the 2D rotation $(x, y, z) \rightarrow (1 - x, y, 1 - z)$, like the mode 1, or in counter-phase, like the mode 2. In the latter case, the antisymmetry with respect to reflection and the 2D rotation leads to the symmetry with respect to the 3D rotation $(x, y, z) \rightarrow (1 - x, 1 - y, 1 - z)$. Note that the definition of the symmetries in Section 2 assumes change of the sign of $\theta = T - (1 - x)$ under either of rotations. Therefore, the mode 1 breaks the 2D rotational symmetry, while the mode 2 breaks the 3D centrosymmetry.

Seeking for an explanation of this instability, we notice the complicated structure of similar flows in two-dimensional cavities, where, at large Grashof numbers, reverse circulations appear at the horizontal boundaries near the corners (see, e.g., Fig. 2e in [34]). In the regions of these reverse circulations, several changes of sign of the vertical velocity along the $x$-axis is observed. We see these changes also in the profiles of the present three-dimensional solution (Fig. 6). The reverse circulations are difficult to recognize in the 3D velocity field, however, they can be revealed in the isolines of the vector potential $\Psi_y$ shown in Figs. 9 and 10 for the cross-section $y = 0.5$. These reverse circulations increase the curvature of streamlines of the main circulation. As a result, the corner flow can become centrifugally unstable. This assumption is elaborated below. The isosurfaces of the absolute values (amplitudes) of vector potential of the most unstable velocity disturbance are shown in the same figure. All the amplitudes are of comparable order of magnitude, and are located in the areas of reverse circulations, which supports the assumption of centrifugal instability.

As mentioned above, appearance of the reverse circulations increases the isolines curvature in the corner regions, so that centrifugal instability mechanisms can be triggered there. To examine this possibility, we calculated the Rayleigh and Bayly criteria [35], as it was done in [22]. Distributions of these criteria are shown in Fig. 11. The Rayleigh criterion, defined as $\eta = -\partial |\mathbf{r} \times \mathbf{v}|^2 / \partial r$, and the radius $\mathbf{r}$ is taken as the distance and direction from the center of main convective circulation. The latter is assumed to coincide with the maximum of the potential $\Psi_y$ in the center plane $y = 0.5$ [22]. An inviscid instability is expected when $\eta$ changes its sign from positive to negative, which clearly takes place Figs. 11a and 11c in the corner areas we wish to examine.

The Bayley criterion $R$ is calculated as in [22,36] using the equation

$$R = \frac{|\boldsymbol{u}_{2D}|^3 \zeta}{(\nabla \Psi_y) \cdot [(\boldsymbol{u}_{2D} \cdot \nabla) \boldsymbol{u}_{2D}]}, \quad \text{where } \boldsymbol{u}_{2D} = rot \Psi_y, \text{ and } \zeta = (rot \boldsymbol{u}_{2D})_y \quad . \quad (10)$$

The inviscid centrifugal instability is expected if $R$ is negative somewhere in the flow. As is seen from Fig. 11a and 11c, the negative values appear in the corners where we observe the largest values of disturbances. Thus, both criteria indicate on possibility of the centrifugal instability, however, related to inviscid flows only, they do not prove it.

To argue additionally in favor of a possible centrifugal mechanism involved in the onset of the observed instability, we compute and plot divergence free projections of the velocity field on the planes parallel and orthogonal to the diagonal plane connecting the lower left and upper right corners, as it was done in [22] for the lid driven cavity flow. The new coordinate planes are defined by the coordinate rotation

$$x' = (x+z)/2, \quad z' = (x-z)/2 \quad . \quad (11)$$

The results for both most unstable modes are shown in Fig. 12. The argument here is that the centrifugal instability results in the vortices rotating in the planes orthogonal to the initial plane of rotation like, e.g., in the Taylor-Couette flow. In Fig. 9 we really observe such vortical structures rotating in the diagonal plane, while the initial "rotating" motion in the corners takes place in the planes $y = const$.

Basing on the above observations, we offer an explanation of the onset of oscillatory instability and the resulting self-sustained oscillatory process as follows. The instability is triggered by the centrifugal mechanism in the corners $x = z = 0$ and $x = z = 1$. As discussed below, the instability in the opposite corners can develop in phase or in counter phase, giving rise to the two most unstable modes having close critical Grashof numbers. The centrifugally induced vortices (Fig. 9) transport colder and warmer fluid in warmer and colder locations, respectively, thus creating the localized temperature perturbations, as is observed in Fig. 5. Note, that a strong convective mixing by the base flow creates stable stratification in most of the bulk of the flow (Fig. 2), except the isothermal vertical boundaries, so that colder fluid is located below the warmer one almost everywhere. Clearly, this stable stratification slows down and suppresses any instability, including the centrifugal one. At the same time, the localized temperature disturbances create local temperature drop in the two horizontal directions, thus creating local non-potential buoyancy forces. This necessarily leads to the appearance of perturbed vortical motion in all the three coordinate planes. The latter is observed in Figs. 7 and 8. This perturbed vortical motion is also suppressed by the stable stratification in the bulk of the flow. At the same time, the perturbed vortical motion in $y = const$ planes, described by the potential $\Psi'_y$ in Figs. 7 and 8 can speed up the main circulation, thus triggering the centrifugal mechanism again.

*4.2.3. Slightly supercritical oscillatory states*

Fully non-linear time-dependent calculations were carried out additionally at slight sub- and super-criticalities. The purposes of this calculations were (i) to examine whether the instability is subcritical, which can explain different frequencies of oscillations of the most unstable modes and calculated oscillatory states (Table 1), and (ii) to examine whether existence of the two leading eigenmodes can lead to development of different oscillatory states. The time-dependent calculations were done as in [14]. The time dependencies of the total kinetic energy $E_{kin}$ and the Nusselt number $Nu$ at the hot wall were monitored. They are defined as

$$E_{kin} = \frac{1}{2}\int_V (u^2 + v^2 + w^2)dV \qquad (12)$$

$$Nu = \int_0^1 \int_0^1 \left[\frac{\partial T}{\partial x}\right]_{x=0} dydz \qquad (13)$$

The first calculation at $Gr = 4.51 \times 10^7$, which is slightly larger than the critical one, results in finite amplitude single frequency oscillations with the dimensionless frequency $\omega \approx 0.0107$. This frequency was reported in [9,14], as well as in [11] for the second mode, see Table 1. This frequency is larger than the frequencies of the two most unstable modes. There is no evidence of any of the frequencies of two most unstable modes of the linearized problem. An increase of the Grashof number to $Gr = 4.6 \times 10^7$, which is slightly larger than the critical value of the second most unstable mode (Table 1), results in the oscillations with a dominant dimensionless frequency $\approx 0.0167$, while a weak peak in the Fourier spectrum is observed also at the dimensionless frequency $\approx 0.00833$. To get a better insight in the frequency change we again used the developed oscillatory flow at $Gr = 4.51 \times 10^7$, and performed calculations for $Gr = 4.55 \times 10^7$. These calculations resulted in an oscillatory regime with $\omega \approx 0.0107$. Further increase of the Grashof number to $Gr = 4.6 \times 10^7$ resulted in the oscillatory state with the same frequency. Thus, we obtained two distinct oscillatory states, which is illustrated in Fig. 13. Figure 13a shows different time dependencies of the kinetic energy, and Fig. 13b different frequency spectra calculated from the two signals of Fig. 13a. Peaks at the reported frequencies and their first harmonics are clearly seen there. Figures 13b and 13c show phase trajectories plotted in the coordinates $(Nu, E_{kin})$. We observe two different phase trajectories, which are double tori, obviously produced by the main frequencies and their first harmonics. For the following we denote the solution branches with the frequency $\approx 0.01$ and $\approx 0.017$, as branches 1 and 2, respectively.

Both branches of the periodic oscillatory flows were calculated for gradually increasing Grashof number until they became stochastic. It was checked also that the solutions of branch 1 are characterized by the 2D rotational symmetry, while the solutions of branch 2 exhibit 3D centrosymmetry. Amplitudes of the primary harmonics were obtained by applying the Fourier transform, and are reported in Fig. 14. One observes that the amplitudes motonically increase with the Grashof number. The filled squares in Fig. 14 show values of the Grashof number, at which the oscillatory state turned into the steady one, when the Grashof number was gradually decreased. These values appear to be smaller than the critical values of the Grashof number reported in Table 1 and shown by filled circles in Fig. 14. The subcritical character of both bifurcations is clearly seen. These subcriticalities explain, in particular, why frequencies of the calculated oscillatory flows differ from those predicted by the linear stability analysis.

At $Gr = 5 \cdot 10^7$ the first branch remains periodic, while the second branch becomes stochastic, as is illustrated by phase diagrams in Fig. 15. The phase diagram of the first branch (Fig. 10a) is a closed line, which returns to itself after making several loops, thus becoming a multiple torus formed by the main oscillation frequency and its several harmonics. The phase diagram of the second branch is an unclosed line localized in space and forming a strange attractor.

### 4.3. Secondary oscillatory – steady bifurcation

As it was found in [14], flows exhibiting stochastic oscillations at $Gr > 5 \times 10^7$ become stable and steady when the Grashof number is increased slightly above $7 \times 10^7$ (see Fig. 9 of [14]). Therefore, the next question to study is the following. Do two distinct stochastic states, as described above, arrive to the same steady flow, or one of them arrives to the steady flow, while the other one remains stochastic, regularly oscillating, or arrives to another steady state? The answer to this question is presented in Figs 16 and 18. In Fig. 16a two phase trajectories that start from some point of the two different stochastic solutions at $Gr \approx 7 \times 10^7$ arrive to two different stochastic solutions in their asymptotic states at $Gr \approx 7.05 \times 10^7$. The difference can be easily seen for the different shapes of the attractors, to which the flows arrive at large time. Figure 16b shows what happens with these two phase trajectories when the Grashof number is further increased to $Gr \approx 7.1 \times 10^7$. We observe that both trajectories arrive to the same stable focus. To illustrate additionally that this is the same steady flow, and not two different flows, we plot the corresponding time dependencies in Fig. 17. It is clearly seen there that that the solutions, belonging to either of two branches, arrive to exactly same values of the kinetic energy and Nusselt number.

We discuss further the instability of steady flow at $Gr < 7.1 \times 10^7$ that occurs when the Grashof number is reduced. The calculations show that the leading eigenvalue, which crosses the imaginary axis, is real and single. The instability develops as a transition from the steady state that preserves all the three symmetries to a stochastically oscillating state, in which all the three symmetries are broken. The vector potentials of the most unstable disturbance are shown in Fig. 18. Note, that in this case of the monotonic instability, the disturbance is a real function. The maximal and minimal values of the potentials indicate on a slight beak of symmetry, so that the symmetry breaking observed in the unsteady flow is a result of the instability, and not a non-linear effect.

Patterns of the vector potential disturbances (Fig. 18) are located in the same areas as for the primary bifurcation (see Fig. 8). In spite that the disturbances structures are not same, they exhibit certain similarity, so that we can perform the same analysis and recall same arguments as was done for the primary bifurcation in Section 4.2.2. Thus, we conclude that also this instability does not develop due to the boundary layers adjacent to the isothermal vertical walls. Furthermore, similarity of the most unstable disturbances leads to an assumption that also in this case the instability sets in owing to centrifugal mechanisms. This assumption is supported by the Rayleigh and Bayly criteria (not shown in figures), which appear to be similar to those depicted in Fig. 11. Finally, we plot the perturbation potential $\Psi'_z$ in the rotated coordinate system, as it was done in Fig. 12. The result is shown in Fig. 19, where we again observe the vortices located in the upper left and lower right corners and rotating in the diagonal plane. As in the case of primary bifurcation, this indicates on the centrifugal instability mechanism.

To explain why the oscillatory flow stabilizes with the increase of the Grashof number, we recall again that at large Grashof numbers the temperature in the bulk of the flow attains a stable stratification, so that colder fluid is located below and warmer fluid above. This is illustrated additionally in cross-sections of Fig. 18c. Therefore, with the increase of the Grashof number the stable stratification becomes steeper, which, at some point, stabilizes the flow. When the Grashof number is decreased, the stabilized steady flow becomes unstable owing to a centrifugal mechanism similar to one that triggers the primary bifurcation. The apparent difference is that this secondary inverse bifurcation cannot be interpreted as a Hopf one, because it sets in due to a single real fully non-symmetric eigenmode.

4.4 The third steady – oscillatory transition

The steady flow, whose stability is reinstated at $Gr \approx 7.1 \cdot 10^7$, preserves all the three symmetries until it becomes oscillatory unstable at $Gr \approx 2.8 \cdot 10^8$. The corresponding most unstable perturbation also exhibits all the three symmetries, which is quite unexpected. Isosurfaces of the temperature disturbance are shown in Fig. 20, and of the divergence free velocity potentials in Fig. 21. The corresponding oscillations can be seen better in the supplied animation files (Animation5 and Animation6).

We observe that the disturbances appear in the upper left and lower right corners. At the first glance these disturbances can be interpreted as rolls elongated in *y*-direction and rotating in the $(x, z)$ planes. However, as is seen from Fig. 21, intensity of the vortical motions in the two other coordinate planes is of the same order. This means that the instability is entirely three-dimensional and cannot be explained as a modification of a known two-dimensional one. Thus, we can argue that this is not instability of the boundary layers [9] because the latter is likely to be two-dimensional-like, owing to the Squire transformation, which can be applied to isothermal, as well as non-isothermal boundary layers. Moreover, the boundary layers are expected to become unstable in the central part of the vertical boundaries, where the shear is larger (see Fig 22), while the observed instability starts in the corners, is advected along the vertical boundaries, and dissipates well before reaching the boundaries middle regions.

We examine the temperature distribution in the upper left and lower right corners (Fig. 3) and find that there is no unstable stratification there. Moreover, the horizontal change of the temperature in these corners appears to be weaker than in the opposite ones. Thus, the onset of instability cannot be attributed to any convective mechanism.

The reverse circulations discussed in Section 4.2 for $Gr = 4.6 \times 10^7$ are observed also at noticeably larger Grashof number, $Gr = 2.96 \times 10^8$, discussed here. They are easily revealed in the vertical velocity profiles that several times change sign along the *x*-direction (Fig. 22). As in the case of the first bifurcation, the reverse circulations are better seen in the isolines of the vector potential $\Psi_y$ shown in Fig. 23 for the cross-section $y = 0.5$ together with the 3D isosurfaces of the absolute values (amplitudes) of vector potential of the most unstable velocity. As in the case of primary bifurcation, all the amplitudes are of comparable order of magnitude, and are located in the areas of reverse circulations.

The Rayleigh and Bayly criteria for the steady flow at $Gr = 2.96 \times 10^8$ are shown in Fig. 24. The patterns of both criteria are quite similar to those plotted in Fig. 9. Again, it is clearly seen that the Rayleigh criterion $\eta$ changes its sign from positive to negative in the corner areas, where the Bayly criterion $R$ attains the negative values. As above, both criteria indicate on possibility of the centrifugal instability.

Following [22] and Section 4.2.2, we calculate the velocity potentials in the rotated coordinate system defined by Eq. (11). Figure 25 shows a snapshot of the potential $\Psi'_z$ in the same way as it was done in Fig. 9. This figure allows one to see vortices located in the diagonal plane shown in the figure, as is expected in the case of centrifugal instability [22]. Amplitudes of the two other velocity potentials are smaller, so that the vortical motion shown in Fig. 25 is dominating. The animation files illustrating this instability also show rotational motion of the perturbation structures in the corners.

As discussed above, if the instability takes place owing to the centrifugal instability mechanism, it has to be the similar in both upper left and lower right corners, so that the perturbation patterns are symmetric or antisymmetric with respect to the 2D rotation (8). Same can be said about the reflection symmetry (7). It is noteworthy that the most unstable disturbance in this case preserves both symmetries. The 3D rotation symmetry, being a superposition of the two former symmetries, is also necessarily preserved. Thus, slightly supercritical oscillatory flows can be expected to preserve all the three symmetries. The oscillatory flows calculated in [14] Gelfgat (2017) for $Gr = 2.9 \times 10^8$ and $3 \times 10^8$ do preserve all the three symmetries. At the same time, existence of the oscillatory flow at $Gr = 2.9 \times 10^8$, which is slightly smaller than the critical one, indicates on the possible subcriticality of this transition.

## 5. Concluding remarks

We revisited the problem of linear stability of 3D buoyancy convection flow in a laterally heated cubical cavity with perfectly insulated horizontal and spanwise boundaries. This problem was considered in several studies by the time-dependent calculations [9-14], where primary steady – oscillatory transition was mainly addressed. The reported results exhibited a noticeable scatter in values of the critical Grashof number, as well as in the frequency of appearing oscillations. Time-dependent calculations of [11] reported two different eigenmodes with quite close critical Grashof (Rayleigh, $Ra = GrPr$) numbers, however this issue since then was never addressed. Later time-dependent calculations of [14] argued that with the increase of the Grashof number, the primary oscillatory instability is followed by flow stabilization, and another steady – oscillatory transition takes place at a much larger Grashof number.

In this study we applied linear stability analysis to the problem and arrived to converged values of the critical Grashof number and critical frequency, which reported and compared with the previous studies in Table 1. Furthermore, we confirmed results of [11] by computing the two distinct eigenmodes. We discussed their symmetries, and argued on the physical reasons that may cause instability due to either of the modes. Applying full 3D time-dependent calculations again, we have shown that steady – oscillatory transitions due to either of modes are subcritical. We also confirmed existence of the chain of bifurcations, so that primary steady – oscillatory transition takes place at $Gr \approx 4.5 \times 10^7$ then the stability reinstates at $Gr \approx 7 \times 10^7$, and the flow becomes oscillatory unstable again at $Gr \approx 2.8 \times 10^8$.

It was shown that time development of the two distinct most unstable eigenmodes result in two different oscillatory solutions that at small supercriticalities preserve the modes symmetries, and with further increase of the Grashof number break all the symmetries and become stochastic. However, when the Grashof number approaches value of $7 \times 10^7$, they arrive to a single steady state, which exhibits all the initial symmetries, and is a stable focus in the phase space.

To gain some more insight into the physical mechanisms that trigger the bifurcations, we analyzed the patterns of flows at the critical Grashof numbers, and patterns of the most unstable disturbances. We argue that the main destabilizing mechanism is centrifugal, which is stabilized by the stable temperature stratification that develops in the bulk of the flow. We have argued that reverse circulations appearing at the horizontal boundaries increase curvature of the streamlines of the main convective circulations, which makes the centrifugal destabilizing mechanism stronger. These three factors appear to be present in all the three transitions.

**ACKNOWLEDGMENTS**

This research was supported by Israel Science Foundation (ISF) grant No 415/18 and was enabled in part by support provided by WestGrid (www.westgrid.ca) and Compute Canada (www.computecanada.ca).

**Figure captions**

Figure 1. Visualization of a subcritical steady 3D flow at $Pr = 0.71$, $Gr = 4.4 \times 10^7$ by isotherms (a) and divergence-free projections of the velocity field on the coordinate planes (b)-(e). The projected velocity fields are depicted by vectors. Isosurfaces of the velocity potentials, to which the projected velocities fields are tangent, are shown by colors. The minimal and maximal values of the potentials are $\pm 0.00205$, $(-0.00982, 0.000226)$, and $\pm 0.00339$ for $\Psi_x$, $\Psi_y$, and $\Psi_z$, respectively. The isosurfaces are plotted for the levels: $-0.0075$ (b) and $-0.002$, $-0.008$, and $-0.009$ (c) for $\Psi_y$; $\pm 0.0008$ for $\Psi_x$ (d); and $\pm 0.001$ for $\Psi_z$ (e).

Figure 2. Visualization of a subcritical steady 3D flow at $Pr = 0.71$, $Gr = 7.2 \times 10^7$ by isotherms (a) and divergence-free projections of the velocity field on the coordinate planes (b)-(e). The projected velocity fields are depicted by vectors. Isosurfaces of the velocity potentials, to which the projected velocities fields are tangent, are shown by colors. The minimal and maximal values of the potentials are $\pm 0.00200$, $(-0.00878, 0.000226)$, and $\pm 0.00308$ for $\Psi_x$, $\Psi_y$, and $\Psi_z$, respectively. The isosurfaces are plotted for the levels: $-0.0006$ (b) and $-0.002$, $-0.0065$, and $-0.008$ (c) for $\Psi_y$; $\pm 0.0008$ for $\Psi_x$ (d); and $\pm 0.001$ for $\Psi^{(z)}$ (e).

Figure 3. Visualization of a subcritical steady 3D flow at $Pr = 0.71$, $Gr = 2.9 \times 10^8$ by isotherms (a) and divergence-free projections of the velocity field on the coordinate planes (b)-(e). The projected velocity fields are depicted by vectors. Isosurfaces of the velocity potentials, to which the projected velocities fields are tangent, are shown by colors. The minimal and maximal values of the potentials are $\pm 0.00174$, $(-0.00642, 0.000838)$, and $\pm 0.00247$ for $\Psi_x$, $\Psi_y$, and $\Psi_z$, respectively. The isosurfaces are plotted for the levels: $-0.004$ (b) and $-0.0025$, $-0.0045$, and $-0.0055$ (c) for $\Psi_y$; $\pm 0.0009$ for $\Psi_x$ (d); and $\pm 0.0012$ for $\Psi_z$ (e).

Figure 4. Amplitude of two modes of the temperature perturbation of the primary bifurcation. (a), (c) – isosurfaces; (b), (d) amplitude of the temperature perturbation in several $y$ = const planes of the frames (a) and (c). Maximal amplitude values are 0.00585 and 0.00474 for the first and second modes, respectively. (e) Amplitude of the slightly supercritical oscillatory 2D flow. (f) Amplitude of the most unstable mode of the temperature perturbation in the 2D AA case, $Gr_{cr}^{(2D)} = 2.42 \times 10^8$.

Figure 5. Real and imaginary parts of two modes of temperature perturbation of the primary bifurcation. The isosurfaces are plotted for the levels $\pm 0.001$, while maximal amplitude values are 0.00585 and 0.00474 for the first and second modes, respectively. Animation1, Animation2.

Figure 6. Profiles of the vertical velocity and the temperature in the spanwise midplane $y = 0.5$. In the frames (b) and (d) the profiles in the interval $0 \leq x \leq 0.2$ are zoomed. $Gr = 4.4 \times 10^7$.

Figure 7. Snapshots of the velocity potentials of the most first most unstable perturbation. The minimal and maximal values and the plotted levels are shown in the upper frames. The snapshots at the 1/2 and 3/4 of the time period can be obtained by reversing the colors. The color maps show the temperature field in the characteristic cross-sections. $Gr = 4.4 \cdot 10^7$. See Supplemental material <Animation 3> for evolution of the perturbation potentials over the time period.

Figure 8. Snapshots of the velocity potentials of the second most unstable perturbation. The minimal and maximal values and the plotted levels are shown in the upper frames. The snapshots at the 1/2 and 3/4 of the time period can be obtained by reversing the colors. The color maps show the temperature field in the characteristic cross-sections. $Gr = 4.6 \times 10^7$. See Supplemental material <Animation 4> for evolution of the perturbation potentials over the time period.

Figure 9. Absolute values (amplitudes) of the velocity potentials of the first most unstable perturbation of the primary bifurcation (isosurfaces) and isolines of the base flow velocity potential $\Psi_y$ plotted in the center plane $y = 0.5$ at $Gr_{cr} \approx 4.6 \times 10^7$. Maximal levels of the amplitudes are shown in the graph. The isosurfaces are plotted for the levels (a) 0.028 and 0.1, (b) 0.012 and 0.2, (c) 0.1 and 0.17.

Figure 10. Absolute values (amplitudes) of the velocity potentials of the second most unstable perturbation of the primary bifurcation (isosurfaces) and isolines of the base flow velocity potential $\Psi_y$ plotted in the center plane $y = 0.5$ at $Gr_{cr} \approx 4.5 \times 10^7$. Maximal levels of the amplitudes are shown in the graph. The isosurfaces are plotted for the levels (a) 0.09 and 0.25, (b) 0.1 and 0.16, (c) 0.09 and 0.15.

Figure 11. Isolines of Rayleigh (a and c) and Bayly (b and d) criteria calculated for the base flow at $Gr_{cr} \approx 4.5 \times 10^7$. The level values in the frames (a) and (b) are: $\pm 0.005$ for the Rayleigh and $\pm 20$ for the Bayly criteria. Positive and negative values are shown by the red and blue color, respectively.

Figure 12. Snapshots of the potential $\Psi'_z(x', y, z')$ obtained after rotation of coordinates by Eq. (10) shown by colors and black isolines for the two most unstable modes at $Gr = 4.6 \times 10^7$. The snapshots are shown at the beginning and a half of the oscillations period. The brown isolines illustrate the main convective circulation in the midplane $y = 0.5$ defined by the vector potential of the base flow $\Psi_y$.

Figure 13. Two distinct oscillatory flow states at $Gr = 4.6 \times 10^7$. The first and second branches are shown by the blue and red colors, respectively. (a) Histories of the total kinetic energy. (b) Fourier spectrum of the dependencies shown in frame (a). (c) and (d) Phase diagrams plotted in the coordinates kinetic energy – Nusselt number.

Figure 14. Amplitudes of first harmonics of the two unstable modes versus the Grashof number for small supercriticalities (lines with empty squares). Filled squares show values of the Grashof

number at which time-dependent computation resulted in a steady flow. Filled circles show critical values of the two modes.

Figure 15. Phase diagrams of the two time-dependent solution branches plotted in the coordinates kinetic energy – Nusselt number at $Gr = 5 \times 10^7$.

Figure 16. Phase trajectories in the plane kinetic energy - Nusselt number. (a) For the two branches starting from two different stochastic states at $7.1 \times 10^7$ and arriving to two different stochastic states at $7.05 \times 10^7$. (b) For the two branches starting from two different stochastic states at $7.05 \times 10^7$ (frame a) and arriving to the same fixed point at $7.1 \times 10^7$.

Figure 17. Time evolution of the kinetic energy (a) and the Nusselt number (b) for the first (red) and second (green) solution branch, when the Grashof number is increased from $7.05 \times 10^7$ to $7.1 \times 10^7$.

Figure 18. Perturbation potentials corresponding to the second bifurcation reinstating the stability. $Gr = 7.08 \times 10^7$. The cross-sections in frame (c) show isotherms equally distributed between the values 0.3 and 0.7. The red color corresponds to larger temperature, and the blue color to smaller ones.

Figure 19. Snapshots of the potential $\Psi'_z(x', y, z')$ obtained after rotation of coordinates by Eq. (10) shown by colors and black isolines at $Gr = 7.08 \times 10^7$. The brown isolines illustrate the main convective circulation in the midplane $y = 0.5$ defined by the vector potential of the base flow $\Psi_y$.

Figure 20. Real and imaginary parts of two modes of temperature perturbation of the primary bifurcation. The isosurfaces are plotted for the levels $\pm 0.0002$, while maximal amplitude value is 0.00384. See supplementary material <Pert_Tmpr+velocities_AAAA_bif3.avi>.

Figure 21. Snapshots of the velocity potentials of the unstable perturbation of the third bifurcation at $Gr_{cr} \approx 2.96 \times 10^8$. The minimal and maximal values and the plotted levels are shown in the upper frames. The snapshots at the 1/2 and 3/4 of the time period can be obtained by reversing the colors. The color maps show the temperature field in the characteristic cross-sections. See Supplemental material <Pert_potentials_bif3.avi> for evolution of the perturbation potentials over the time period.

Figure 22. Profiles of the vertical velocity and the temperature in the spanwise midplane $y = 0.5$. In the frames (b) and (d) the profiles in the interval $0 \leq x \leq 0.2$ are zoomed. $Gr_{cr} \approx 2.96 \times 10^8$.

Figure 23. Absolute values of the velocity potentials of the most unstable perturbation of the third bifurcation (isosurfaces) and isolines of the base flow velocity potential $\Psi_y$ plotted in the center

plane $y = 0.5$ at $Gr_{cr} \approx 2.96 \times 10^8$. The isosurfaces are plotted for the levels 0.1 and 0.25. The maximal values of $|\Psi'_x|, |\Psi'_y|$ and $|\Psi'_z|$ are 0.640, 0.747, and 0.468, respectively.

Figure 24. Isolines of Rayleigh (a and c) and Bayly (b and d) criteria calculated for the base flow at $Gr_{cr} \approx 2.96 \times 10^8$. The level values in the frames (a) and (b) are: $\pm 0.03$ for the Rayleigh and $\pm 20$ for the Bayly criteria. Positive and negative values are shown by the red and blue color, respectively.

Figure 25. Snapshot of the potential $\Psi'_z(x', y, z')$ obtained after rotation of coordinates by Eq. (10) shown by colors and black isolines for the most unstable mode at $Gr = 2.96 \times 10^8$. The snapshots are shown at the beginning and a half of the oscillations period. The brown isolines illustrate the main convective circulation in the midplane $y = 0.3$ defined by the vector potential of the base flow $\Psi_y$.

Table 1. Present linear stability results compared with the previous ones obtained by the time-dependent calculations.

| Reference | Method | $Gr_{cr} \cdot 10^{-7}$ | $f_{cr}$ |
|---|---|---|---|
| 1st bifurcation | | | |
| Janssen & Henkes (1995) [9] | $120^3$ finite volume grid | 3.5 ÷ 5.6 | 0.01 |
| Labrosse et al. (1997) [10] | $61^3$ collocation points | 4.79 | 0.00488 |
| De Gassowski et al (2006) [11] | $80^3$ collocation points | 4.46<br>4.52 | 0.00502<br>0.00919 |
| Soucasse et al (2014) [13] | $80^3$ collocation points | 4.4 ÷ 4.9 | 0.0067 |
| Gelfgat (2017) [14] | $256^3$ finite volume grid | 4.58 | 0.0107 |
| Present | $256^3$ finite volume grid, linear stability analysis | 4.504<br>4.517 | 0.00504<br>0.00791 |
| 2nd bifurcation | | | |
| Gelfgat (2017) [14] | $256^3$ finite volume grid | 7.2 | 0 |
| Present | $256^3$ finite volume grid, linear stability analysis | 7.095 | 0 |
| 3rd bifurcation | | | |
| Gelfgat (2017) [14] | $256^3$ finite volume grid | 28.0 | 0.525 |
| Present | $256^3$ finite volume grid, linear stability analysis | 29.59 | 0.531 |

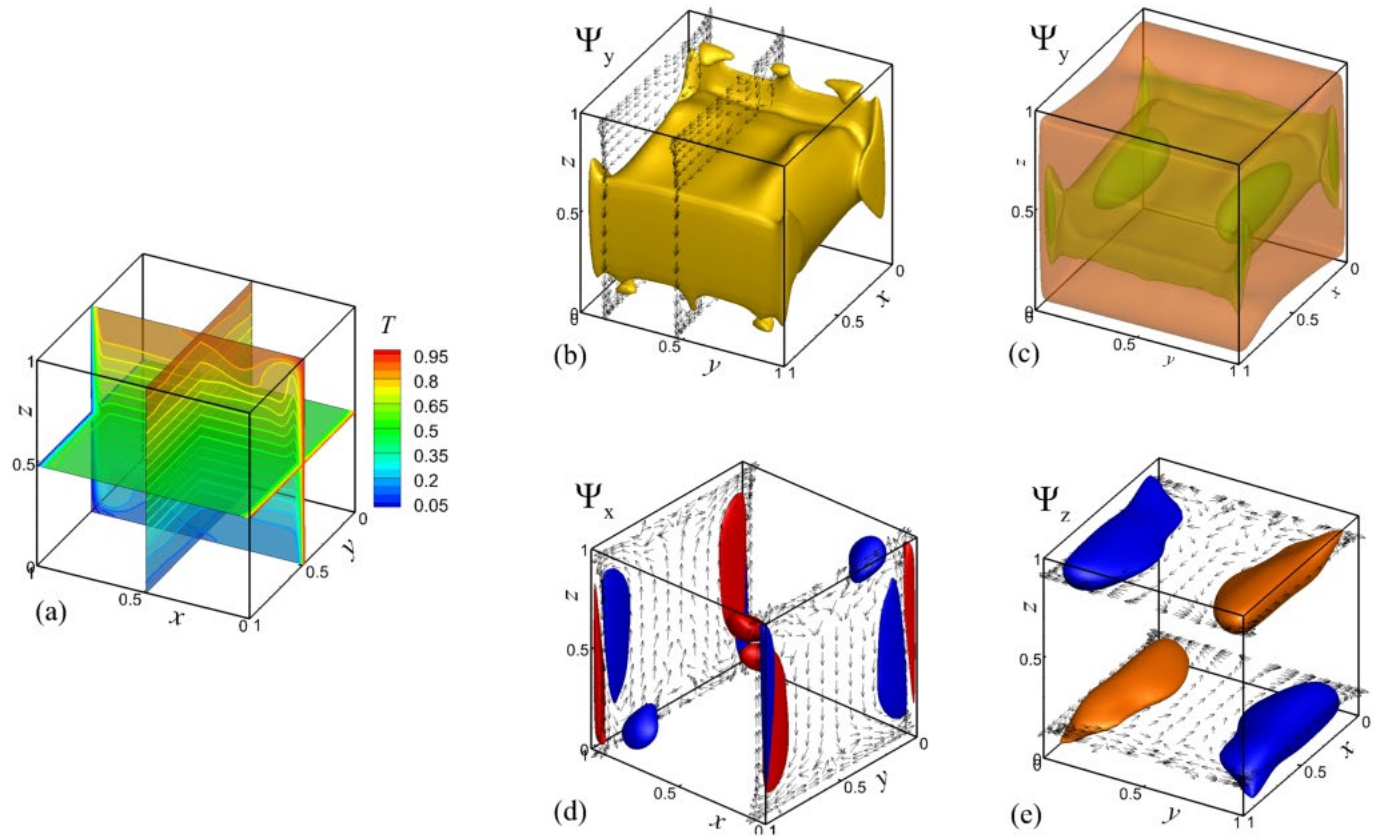

Figure 1. Visualization of a subcritical steady 3D flow at $Pr = 0.71$, $Gr = 4.4 \times 10^7$ by isotherms (a) and divergence-free projections of the velocity field on the coordinate planes (b)-(e). The projected velocity fields are depicted by vectors. Isosurfaces of the velocity potentials, to which the projected velocities fields are tangent, are shown by colors. The minimal and maximal values of the potentials are $\pm 0.00205$, $(-0.00982, 0.000226)$, and $\pm 0.00339$ for $\Psi_x$, $\Psi_y$, and $\Psi_z$, respectively. The isosurfaces are plotted for the levels: $-0.0075$ (b) and $-0.002$, $-0.008$, and $-0.009$ (c) for $\Psi_y$; $\pm 0.0008$ for $\Psi_x$ (d); and $\pm 0.001$ for $\Psi_z$ (e).

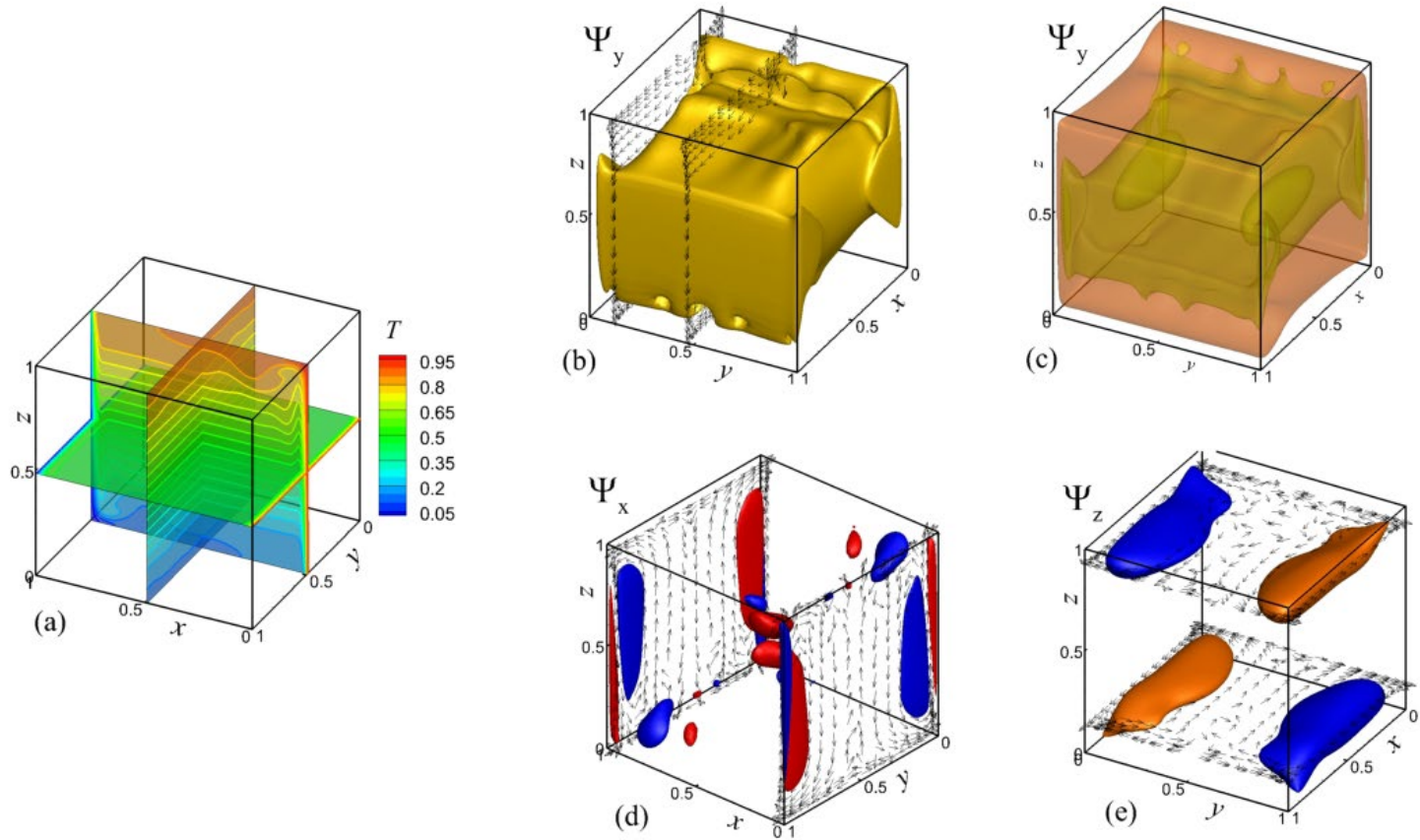

Figure 2. Visualization of a subcritical steady 3D flow at $Pr = 0.71$, $Gr = 7.2 \times 10^7$ by isotherms (a) and divergence-free projections of the velocity field on the coordinate planes (b)-(e). The projected velocity fields are depicted by vectors. Isosurfaces of the velocity potentials, to which the projected velocities fields are tangent, are shown by colors. The minimal and maximal values of the potentials are $\pm 0.00200$, $(-0.00878, 0.000226)$, and $\pm 0.00308$ for $\Psi_x$, $\Psi_y$, and $\Psi_z$, respectively. The isosurfaces are plotted for the levels: $-0.0006$ (b) and $-0.002$, $-0.0065$, and $-0.008$ (c) for $\Psi_y$; $\pm 0.0008$ for $\Psi_x$ (d); and $\pm 0.001$ for $\Psi^{(z)}$ (e).

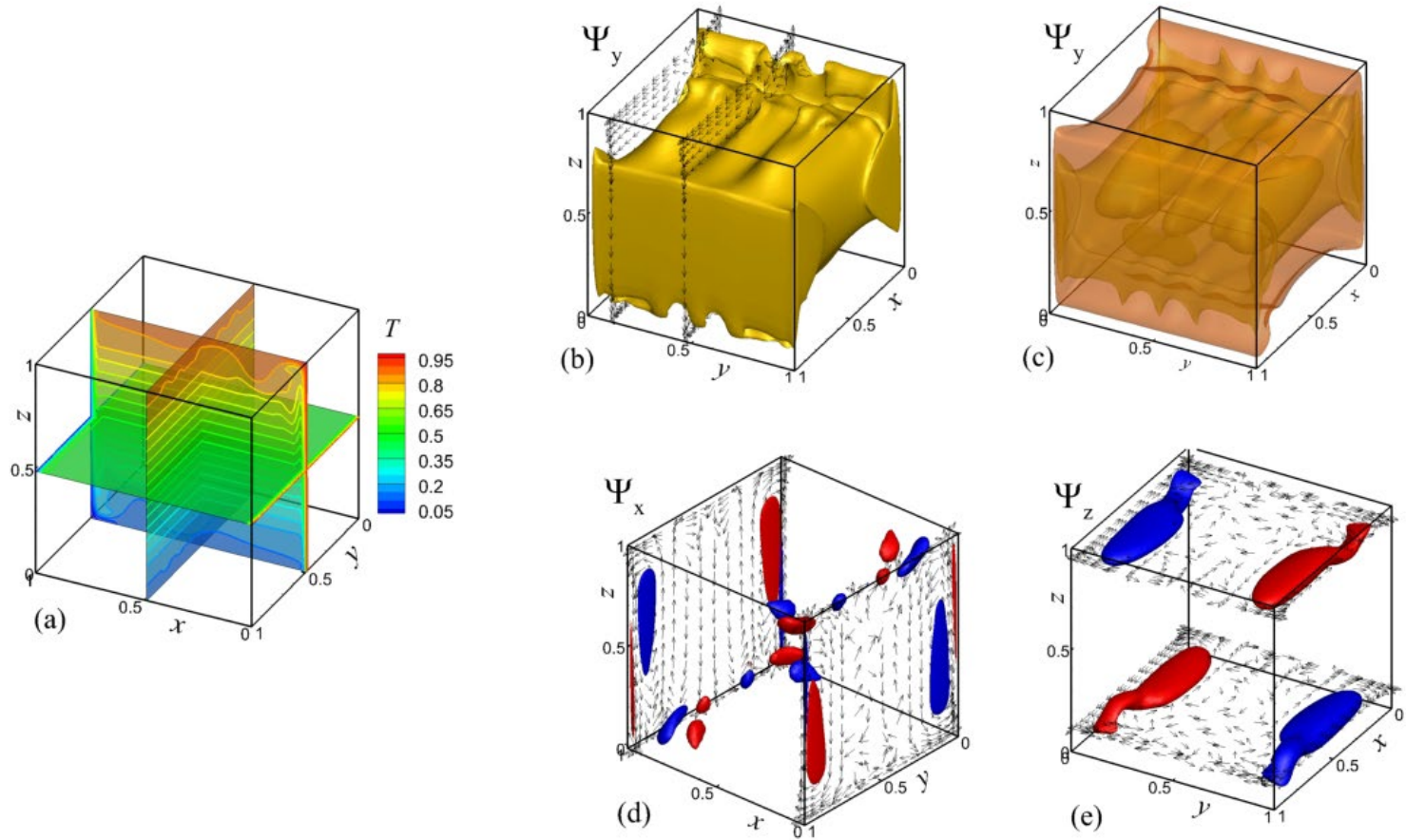

Figure 3. Visualization of a subcritical steady 3D flow at $Pr = 0.71$, $Gr = 2.9 \times 10^8$ by isotherms (a) and divergence-free projections of the velocity field on the coordinate planes (b)-(e). The projected velocity fields are depicted by vectors. Isosurfaces of the velocity potentials, to which the projected velocities fields are tangent, are shown by colors. The minimal and maximal values of the potentials are $\pm 0.00174$, $(-0.00642, 0.000838)$, and $\pm 0.00247$ for $\Psi_x$, $\Psi_y$, and $\Psi_z$, respectively. The isosurfaces are plotted for the levels: $-0.004$ (b) and $-0.0025, -0.0045$, and $-0.0055$ (c) for $\Psi_y$; $\pm 0.0009$ for $\Psi_x$ (d); and $\pm 0.0012$ for $\Psi_z$ (e).

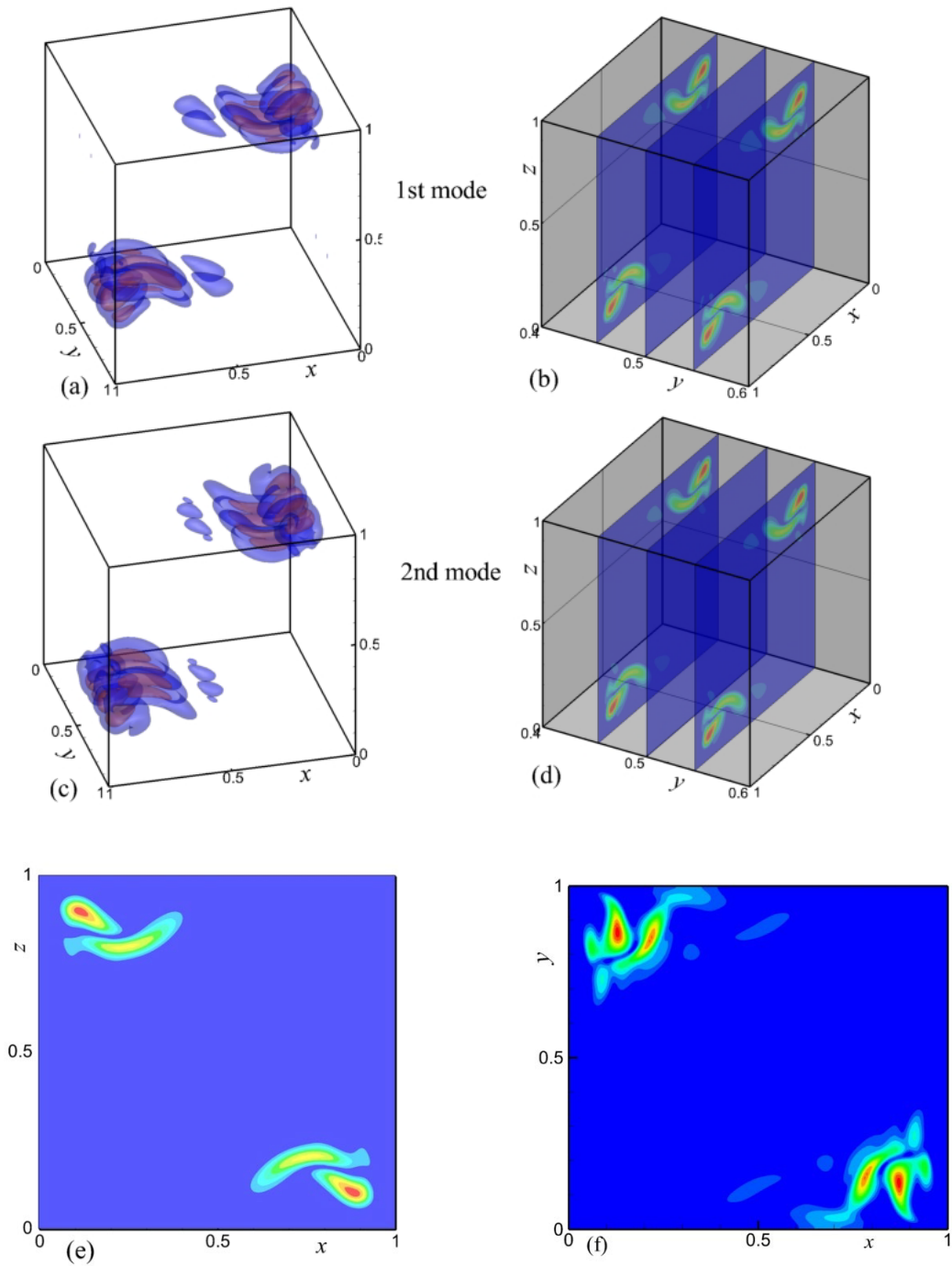

Figure 4. Amplitude of two modes of the temperature perturbation of the primary bifurcation. (a), (c) – isosurfaces; (b), (d) amplitude of the temperature perturbation in several $y$ = const planes of the frames (a) and (c). Maximal amplitude values are 0.00585 and 0.00474 for the first and second modes, respectively. (e) Amplitude of the slightly supercritical oscillatory 2D flow. (f) Amplitude of the most unstable mode of the temperature perturbation in the 2D AA case, $Gr_{cr}^{(2D)} = 2.42 \times 10^8$.

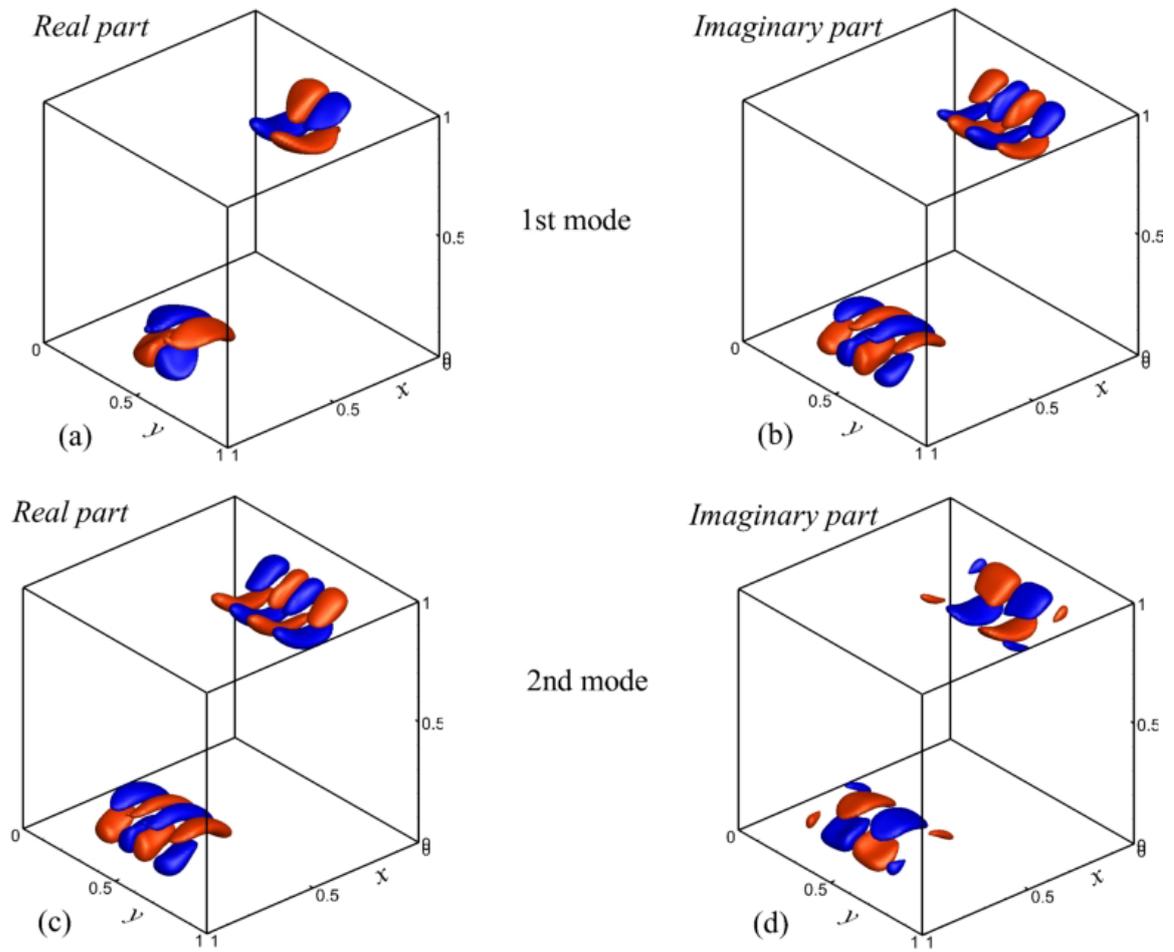

Figure 5. Real and imaginary parts of two modes of temperature perturbation of the primary bifurcation. The isosurfaces are plotted for the levels ±0.001, while maximal amplitude values are 0.00585 and 0.00474 for the first and second modes, respectively. Animation1, Animation2.

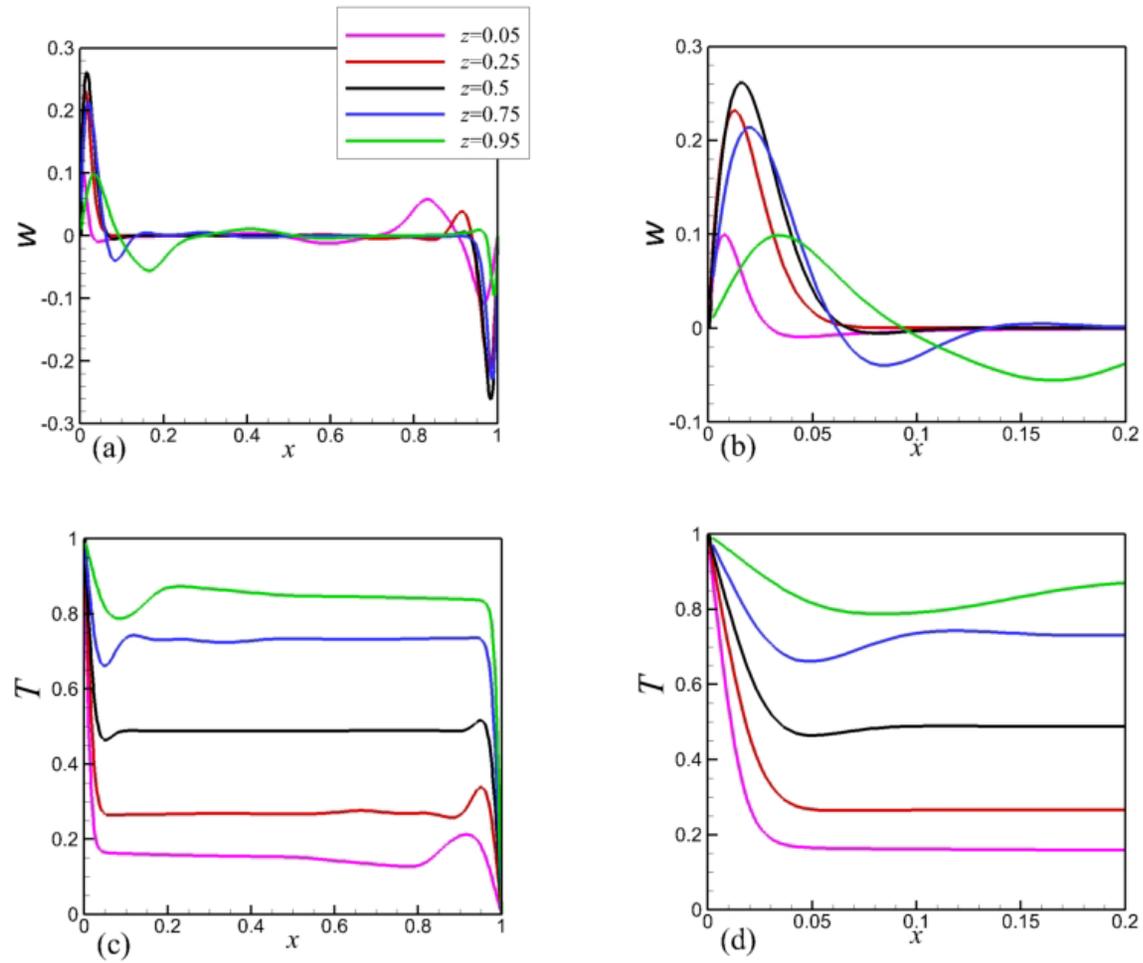

Figure 6. Profiles of the vertical velocity and the temperature in the spanwise midplane $y = 0.5$. In the frames (b) and (d) the profiles in the interval $0 \leq x \leq 0.2$ are zoomed. $Gr = 4.4 \times 10^7$.

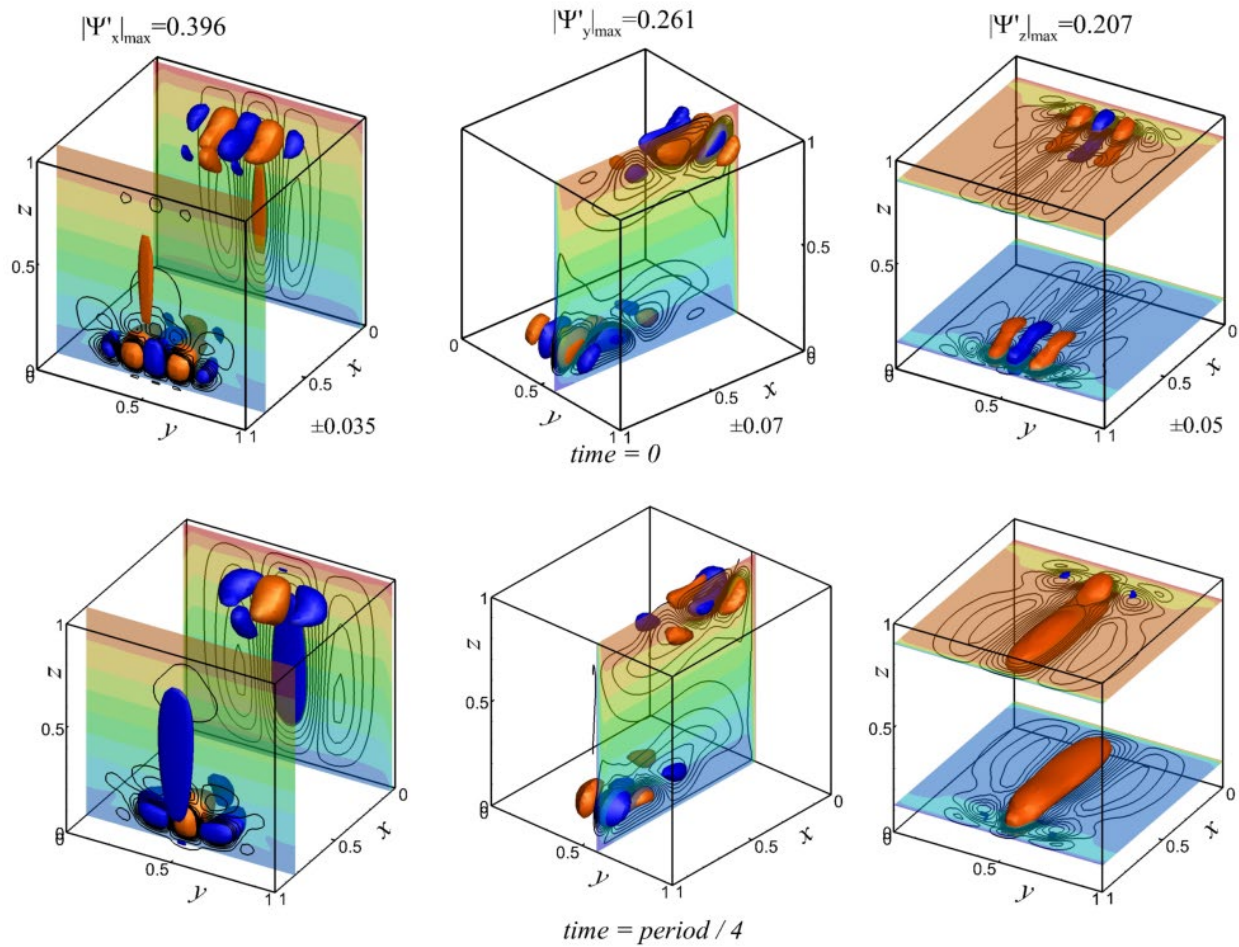

Figure 7. Snapshots of the velocity potentials of the first most unstable perturbation. The minimal and maximal values and the plotted levels are shown in the upper frames. The snapshots at the 1/2 and 3/4 of the time period can be obtained by reversing the colors. The color maps show the temperature field in the characteristic cross-sections. $Gr = 4.5 \times 10^7$. See Supplemental material <Animation 3> for evolution of the perturbation potentials over the time period.

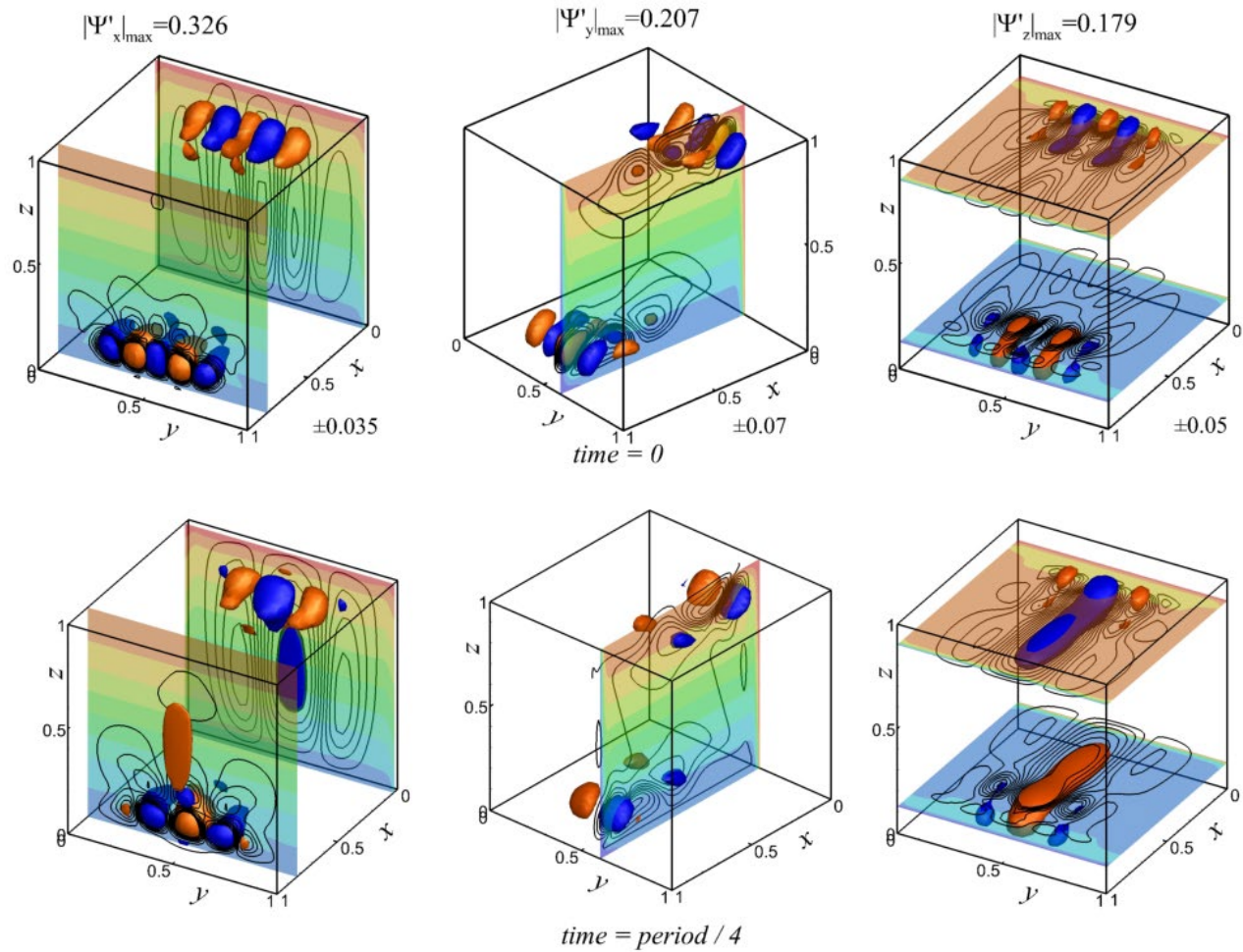

Figure 8. Snapshots of the velocity potentials of the second most unstable perturbation. The minimal and maximal values and the plotted levels are shown in the upper frames. The snapshots at the 1/2 and 3/4 of the time period can be obtained by reversing the colors. The color maps show the temperature field in the characteristic cross-sections. $Gr = 4.5 \times 10^7$. See Supplemental material <Animation 4> for evolution of the perturbation potentials over the time period.

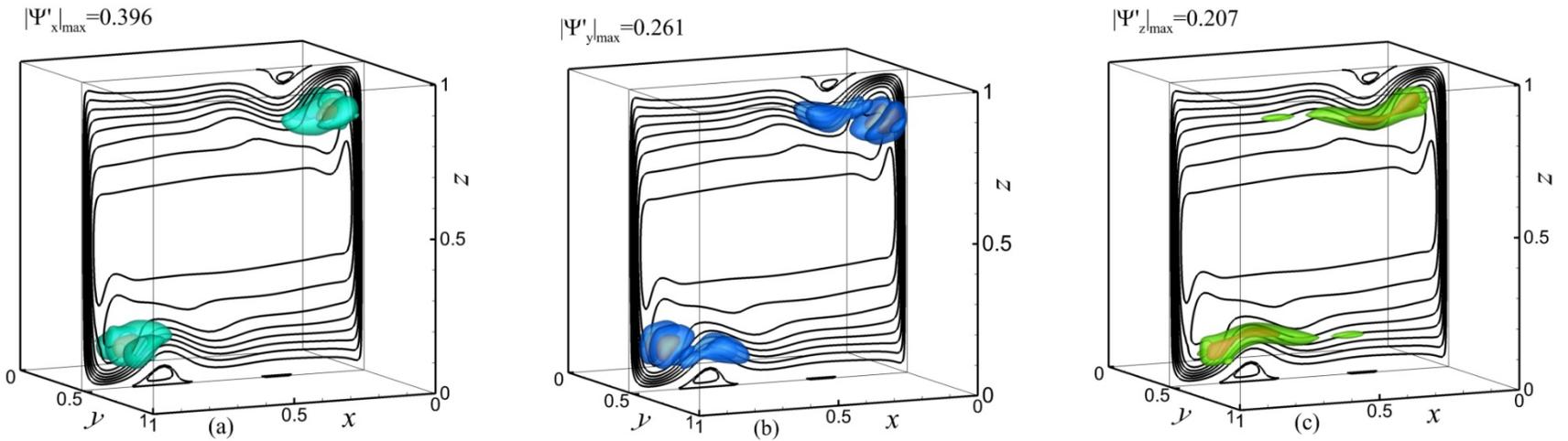

Figure 9. Absolute values (amplitudes) of the velocity potentials of the first most unstable perturbation of the primary bifurcation (isosurfaces) and isolines of the base flow velocity potential $\Psi_y$ plotted in the center plane $y = 0.5$ at $Gr_{cr} \approx 4.5 \times 10^7$. Maximal levels of the amplitudes are shown in the graph. The isosurfaces are plotted for the levels (a) 0.028 and 0.1, (b) 0.012 and 0.2, (c) 0.1 and 0.17.

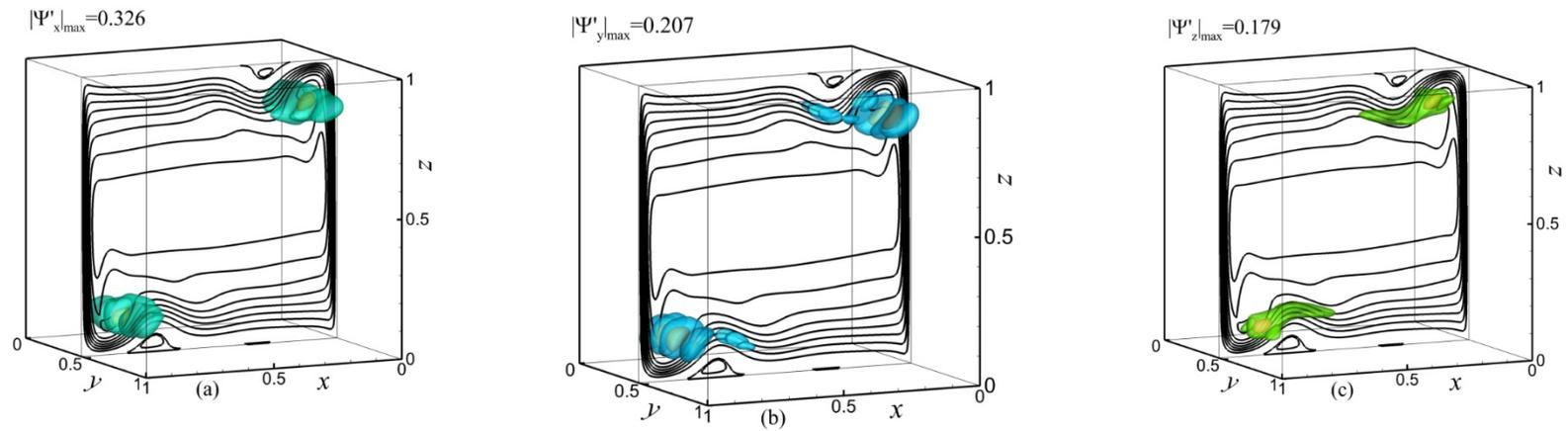

Figure 10. Absolute values (amplitudes) of the velocity potentials of the second most unstable perturbation of the primary bifurcation (isosurfaces) and isolines of the base flow velocity potential $\Psi_y$ plotted in the center plane $y = 0.5$ at $Gr_{cr} \approx 4.5 \times 10^7$. Maximal levels of the amplitudes are shown in the graph. The isosurfaces are plotted for the levels (a) 0.09 and 0.25, (b) 0.1 and 0.16, (c) 0.09 and 0.15.

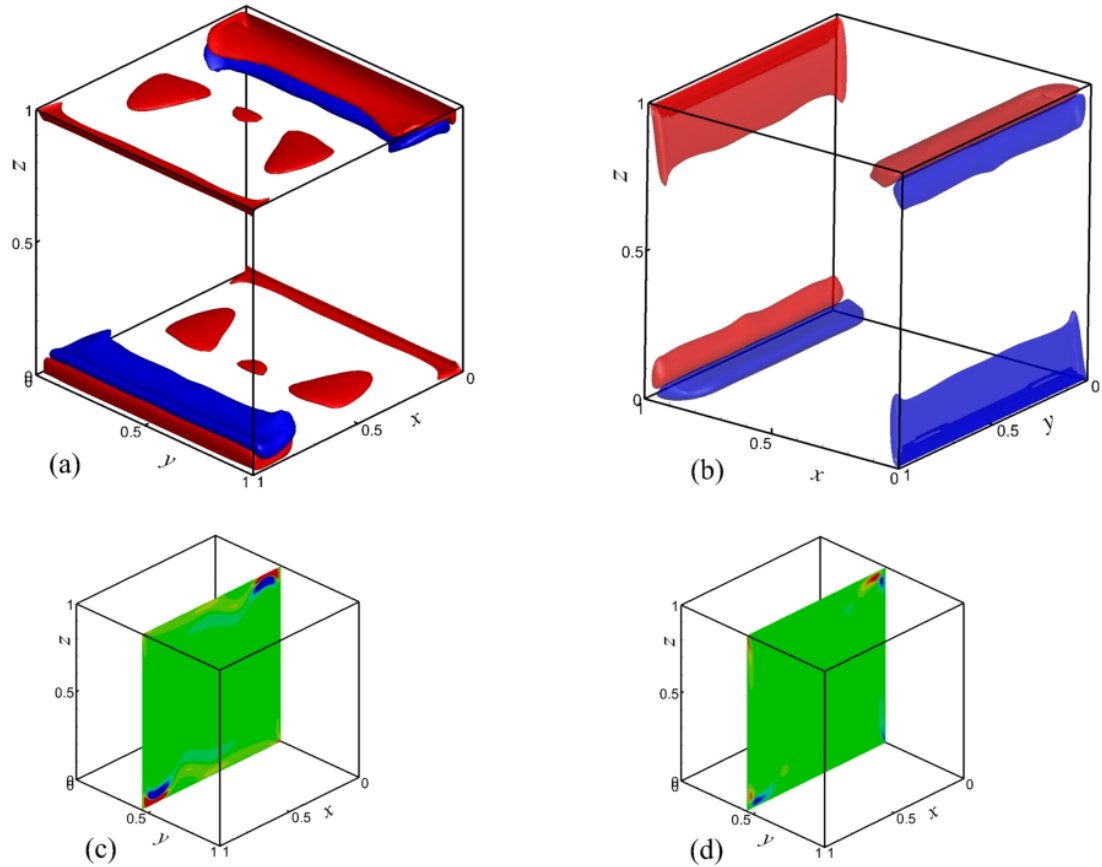

Figure 11. Isolines of Rayleigh (a and c) and Bayly (b and d) criteria calculated for the base flow at $Gr_{cr} \approx 4.5 \times 10^7$. The level values in the frames (a) and (b) are: $\pm 0.005$ for the Rayleigh and $\pm 20$ for the Bayly criteria. Positive and negative values are shown by the red and blue color, respectively.

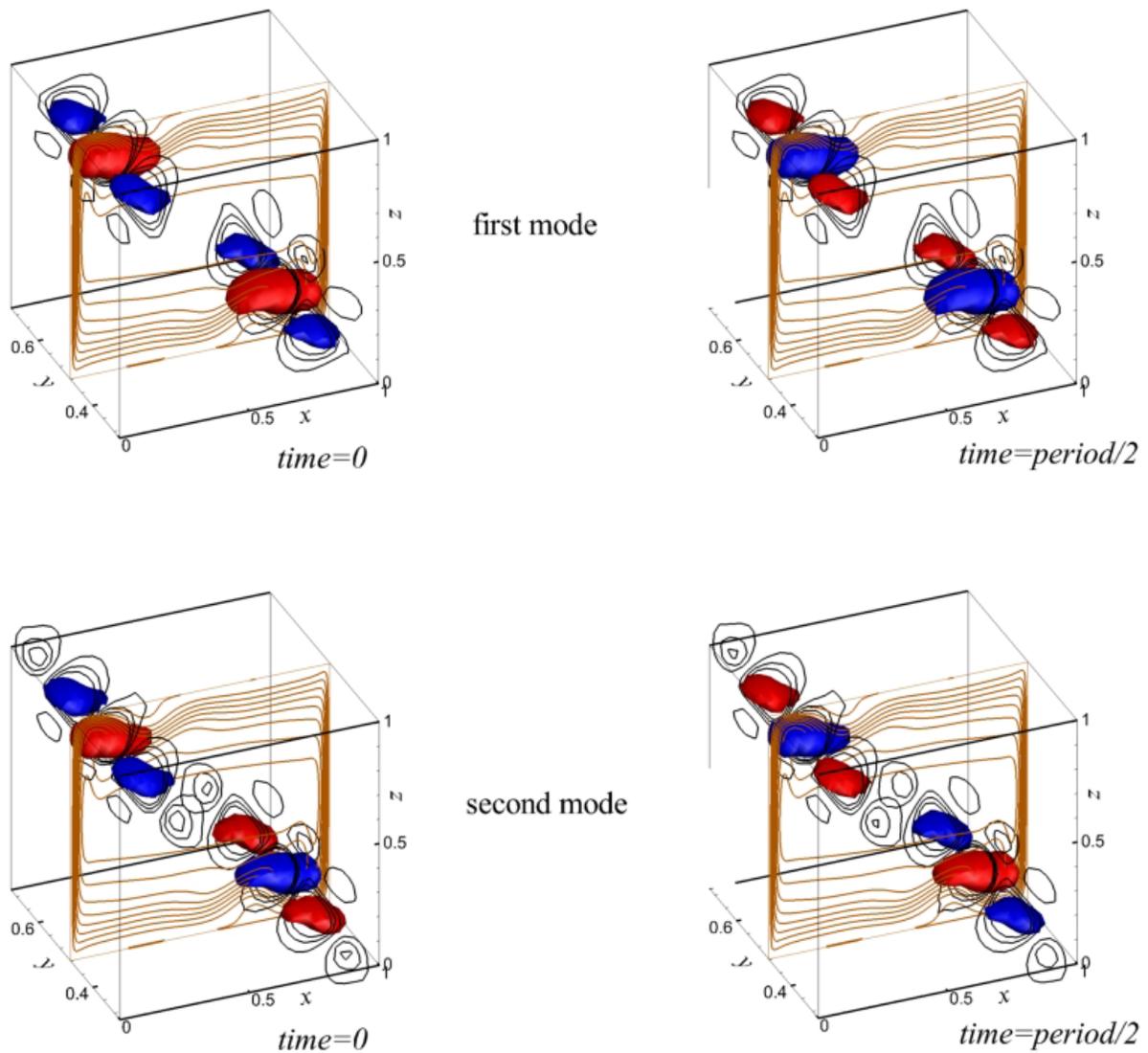

Figure 12. Snapshots of the potential $\Psi'_z(x', y, z')$ obtained after rotation of coordinates by Eq. (10) shown by colors and black isolines for the two most unstable modes at $Gr = 4.6 \times 10^7$. The snapshots are shown at the beginning and a half of the oscillations period. The brown isolines illustrate the main convective circulation in the midplane $y = 0.5$ defined by the vector potential of the base flow $\Psi_y$.

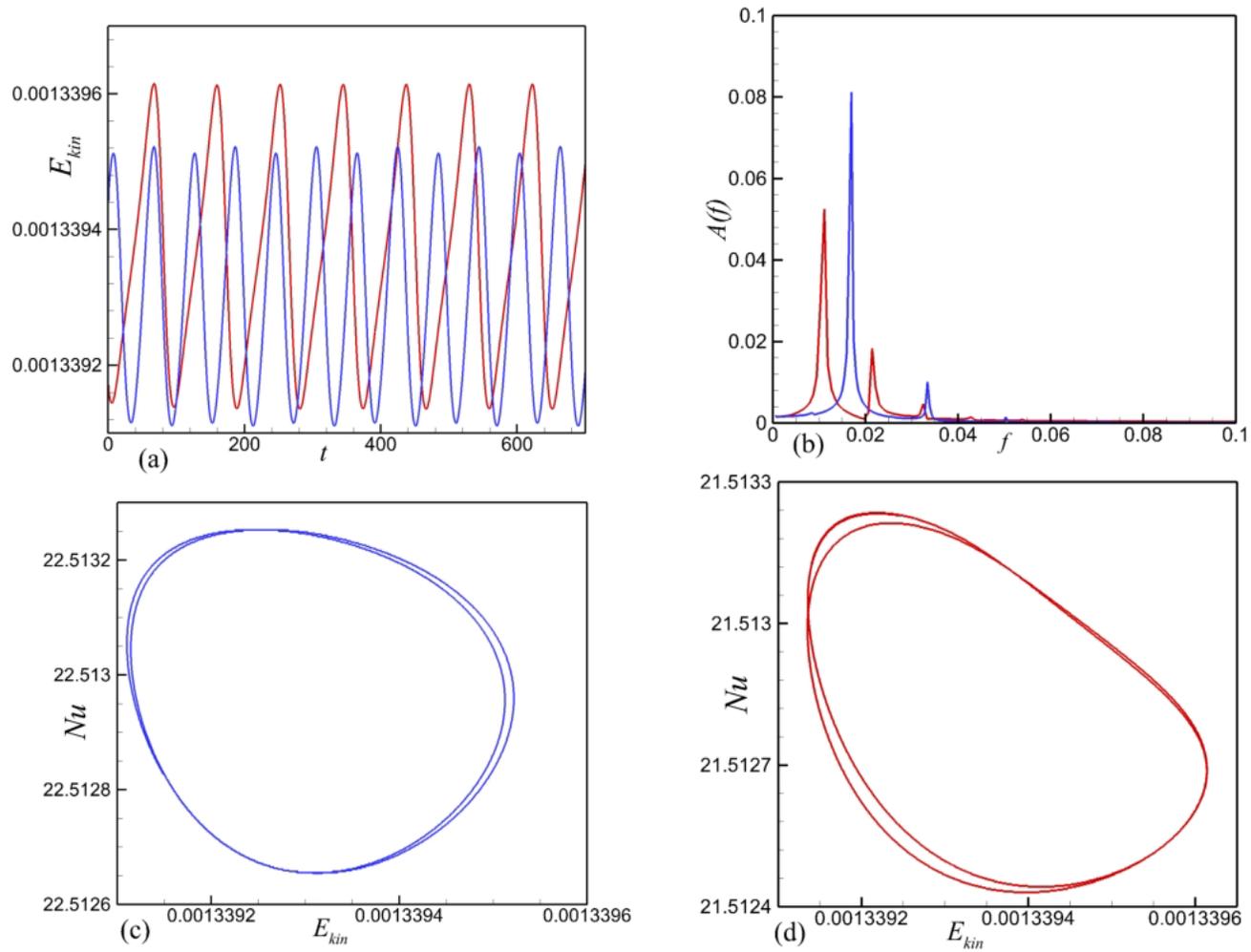

Figure 13. Two distinct oscillatory flow states at $Gr = 4.6 \times 10^7$. The first and second branches are shown by the red and blue colors, respectively. (a) Histories of the total kinetic energy. (b) Fourier spectrum of the dependencies shown in frame (a). (c) and (d) Phase diagrams plotted in the coordinates kinetic energy – Nusselt number.

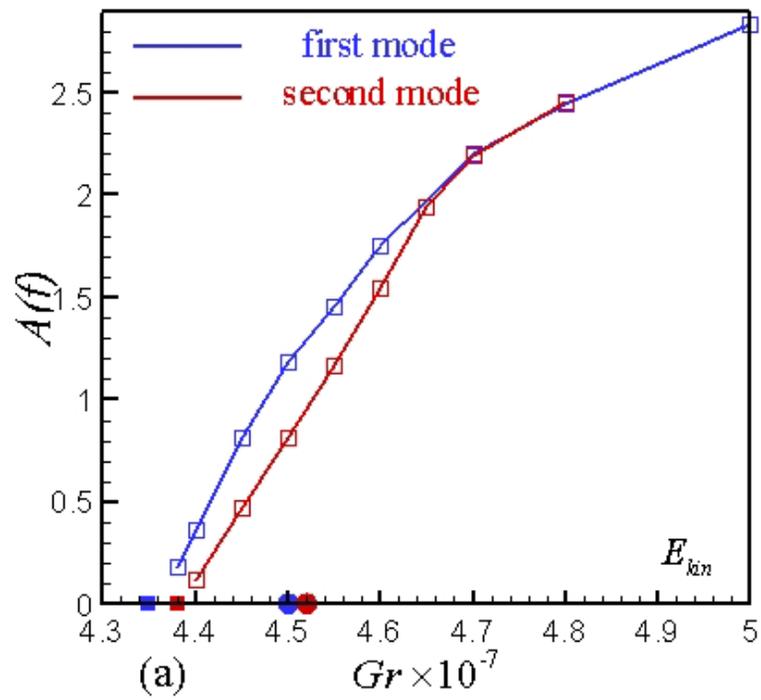 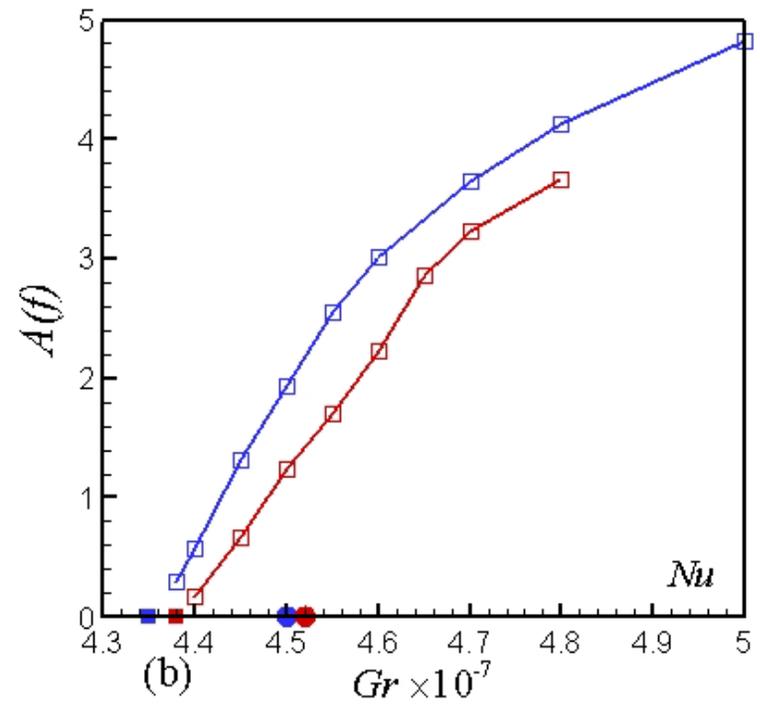

Figure 14. Amplitudes of first harmonics of the two unstable modes versus the Grashof number for small supercriticalities (lines with empty squares). Filled squares show values of the Grashof number at which time-dependent computation resulted in a steady flow. Filled circles show critical values of the two modes.

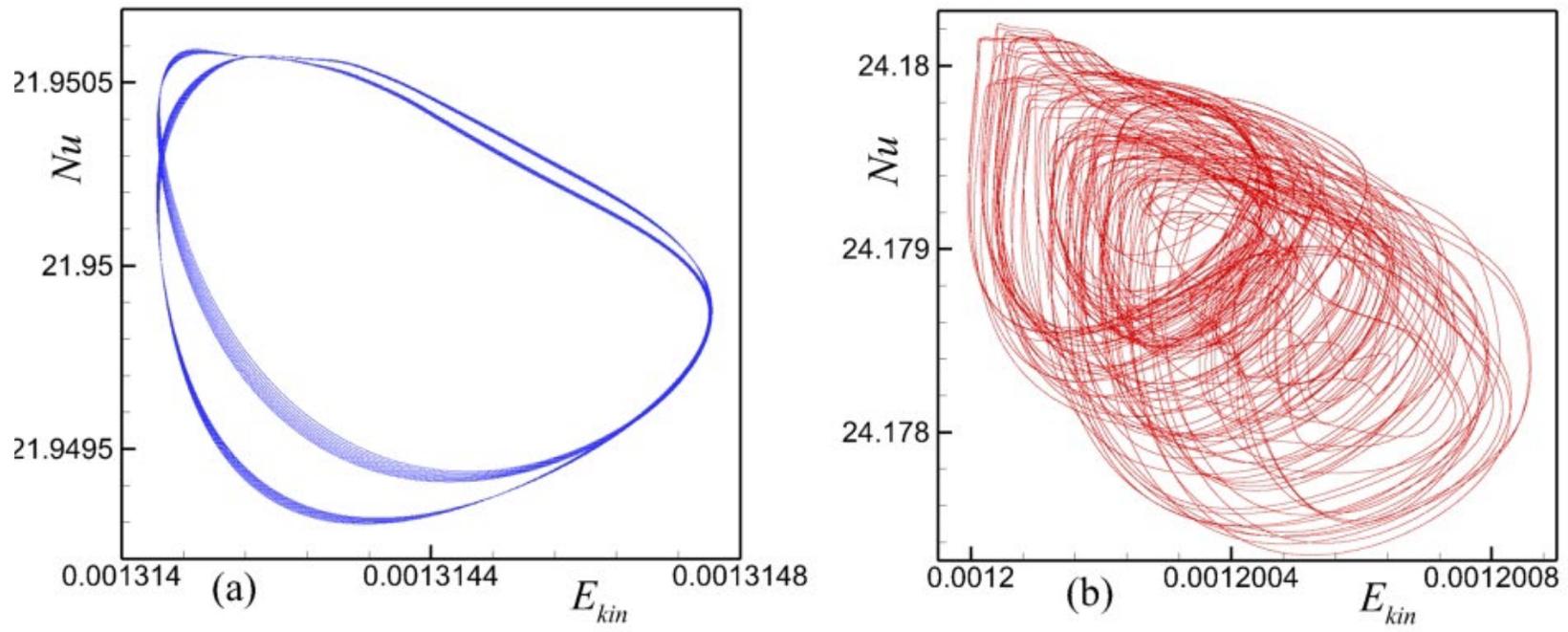

Figure 15. Phase diagrams of the two time-dependent solution branches plotted in the coordinates kinetic energy – Nusselt number at $Gr = 5 \times 10^7$.

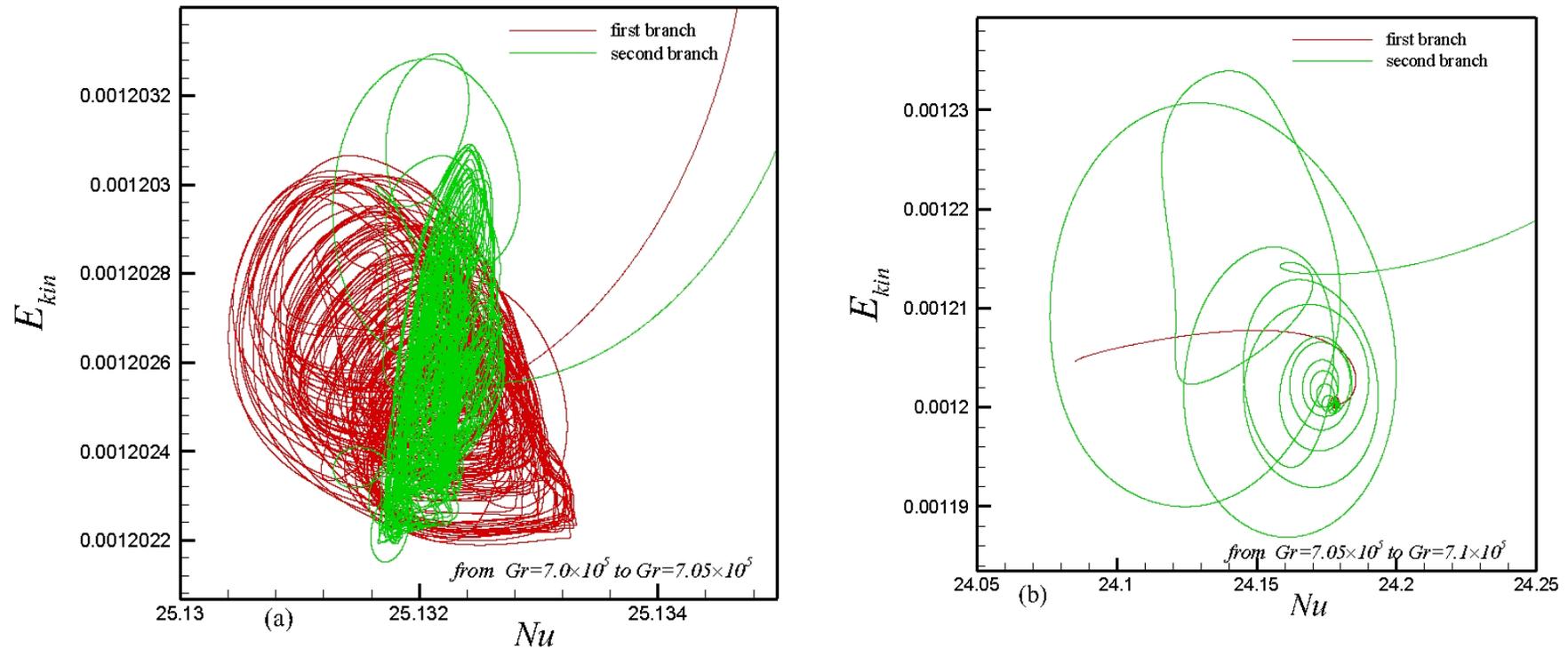

Figure 16. Phase trajectories in the plane kinetic energy - Nusselt number. (a) For the two branches starting from two different stochastic states at $7.1 \times 10^7$ and arriving to two different stochastic states at $7.05 \times 10^7$. (b) For the two branches starting from two different stochastic states at $7.05 \times 10^7$ (frame a) and arriving to the same fixed point at $7.1 \times 10^7$.

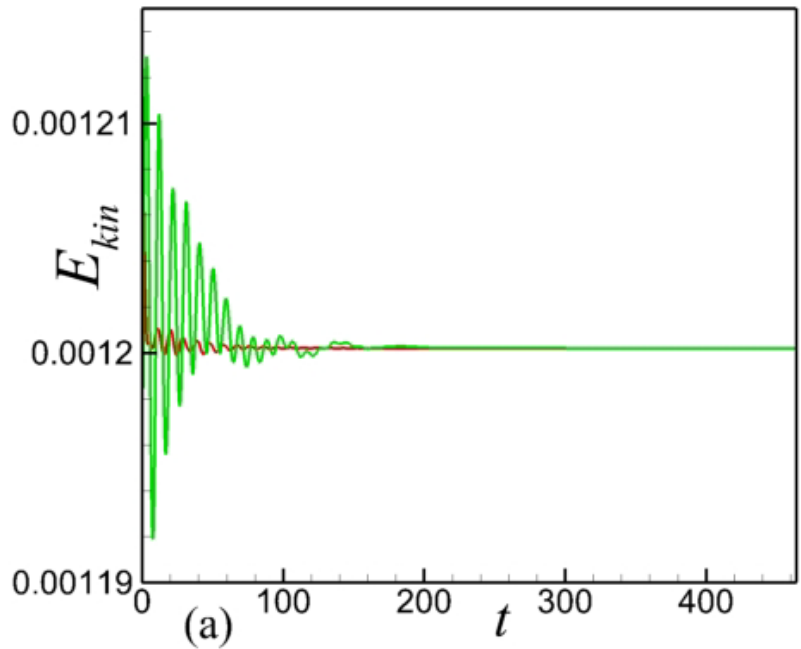 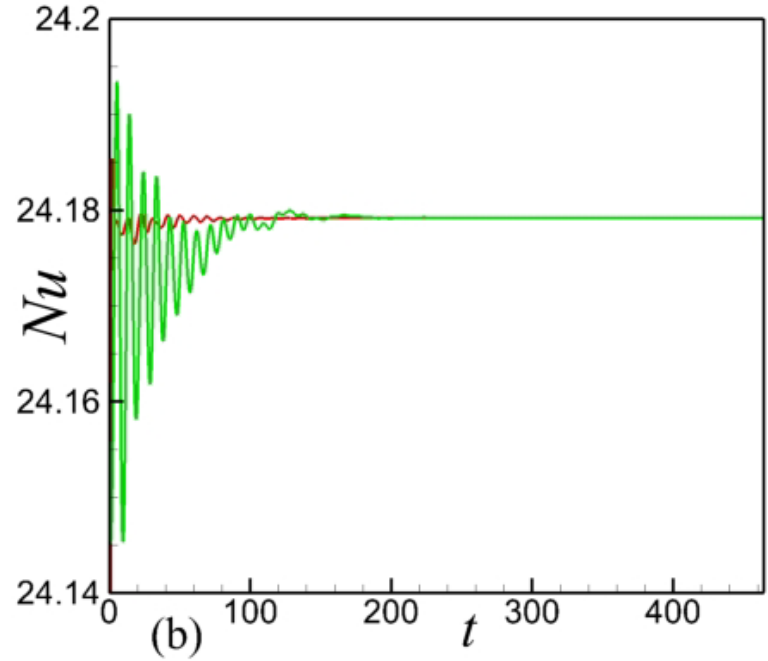

Figure 17. Time evolution of the kinetic energy (a) and the Nusselt number (b) for the first (red) and second (green) solution branch, when the Grashof number is increased from $7.05 \times 10^7$ to $7.1 \times 10^7$.

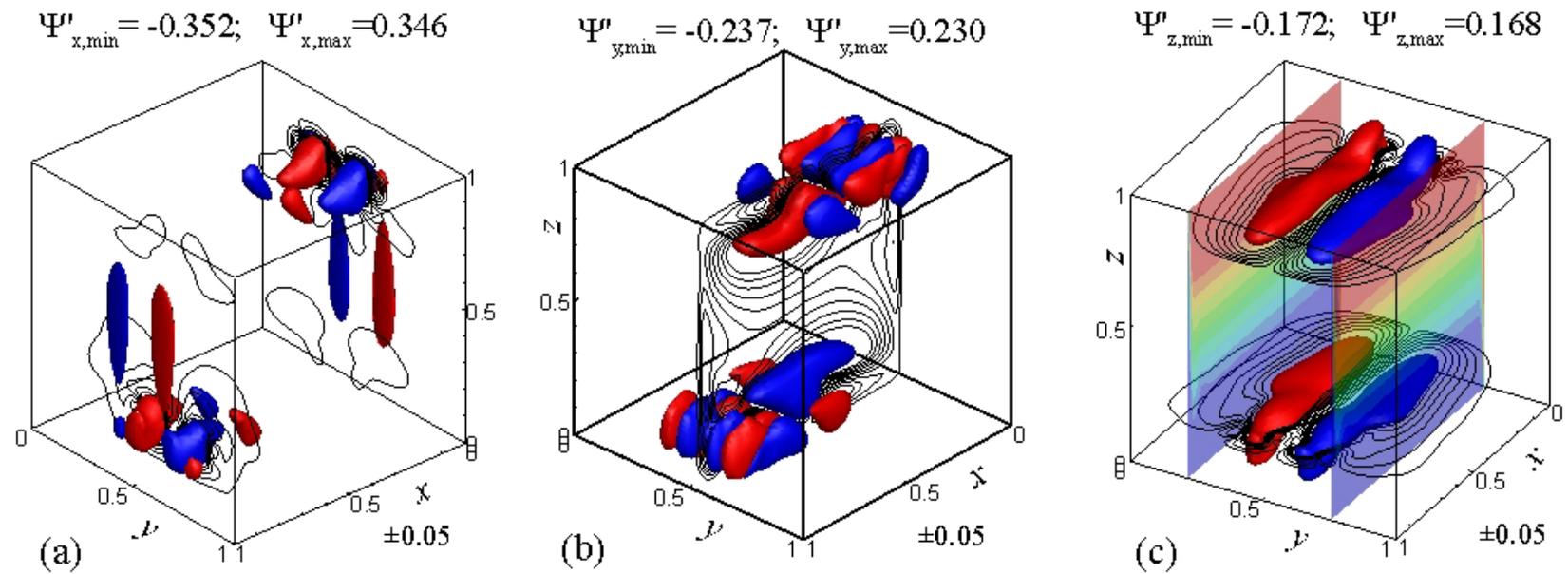

Figure 18. Perturbation potentials corresponding to the second bifurcation reinstating the stability. $Gr = 7.08 \times 10^7$. The cross-sections in frame (c) show isotherms equally distributed between the values 0.3 and 0.7. The red color corresponds to larger temperature, and the blue color to smaller ones.

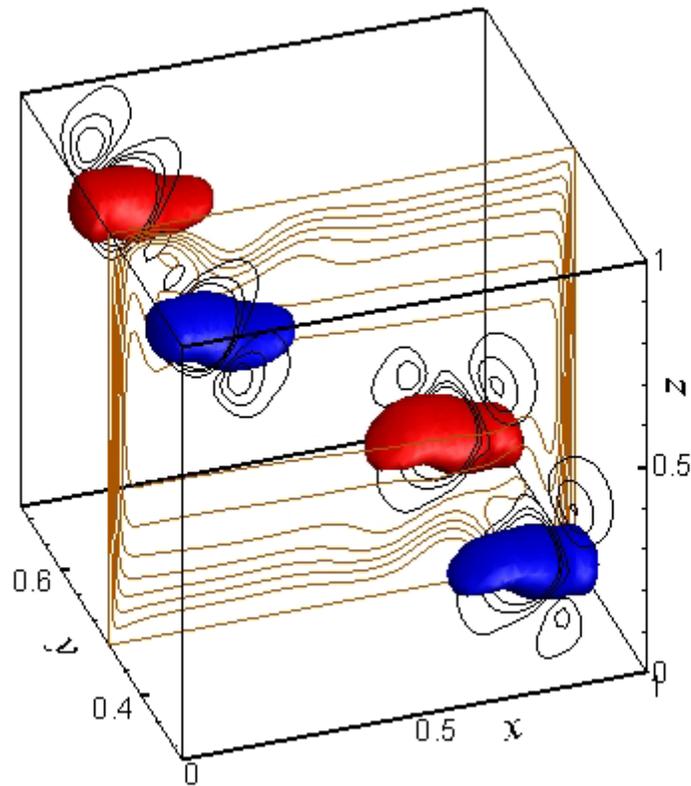

Figure 19. Snapshots of the potential $\Psi'_z(x', y, z')$ obtained after rotation of coordinates by Eq. (10) shown by colors and black isolines at $Gr = 7.08 \times 10^7$. The brown isolines illustrate the main convective circulation in the midplane $y = 0.5$ defined by the vector potential of the base flow $\Psi_y$.

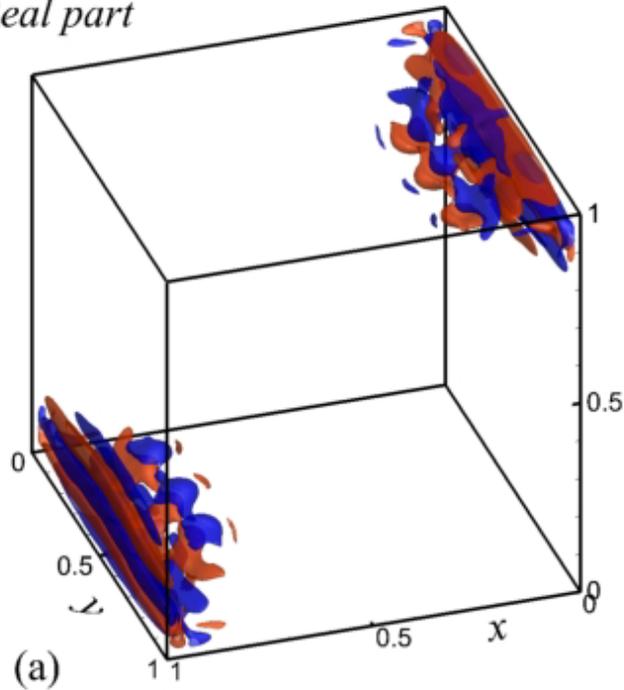 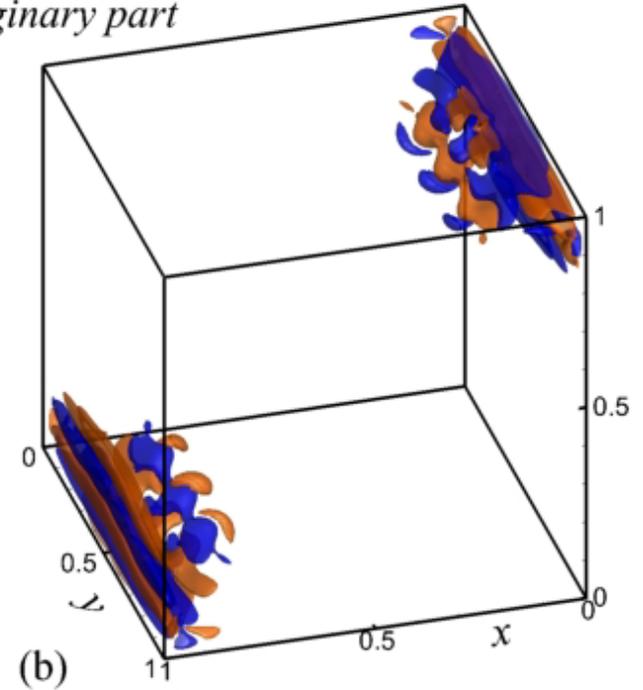

Figure 20. Real and imaginary parts of two modes of temperature perturbation of the primary bifurcation. The isosurfaces are plotted for the levels ±0.0002, while maximal amplitude value is 0.00384. See supplementary material <Pert_Tmpr+velocities_AAAA_bif3.avi>.

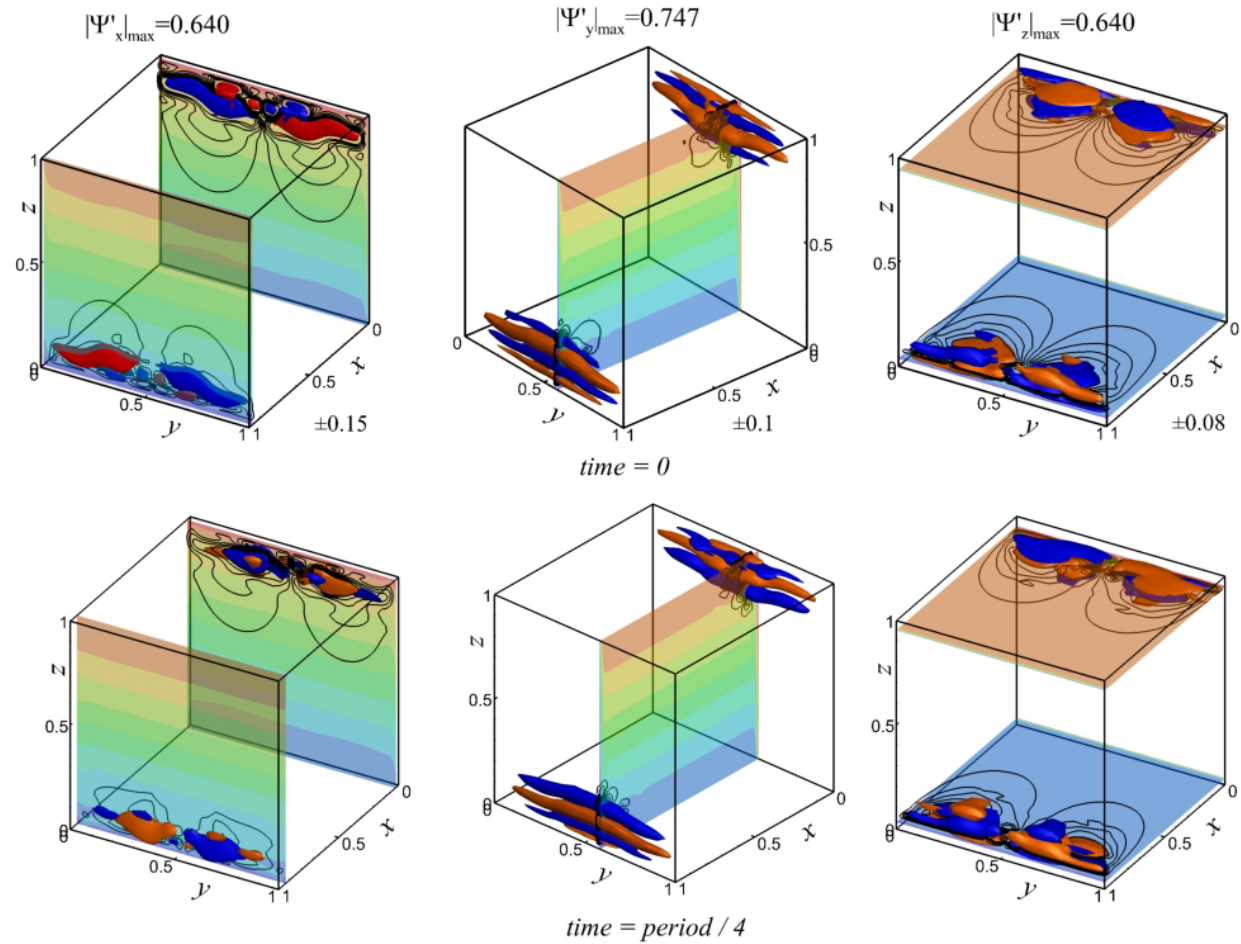

Figure 21. Snapshots of the velocity potentials of the most unstable perturbation of the third bifurcation at $Gr_{cr} \approx 2.96 \times 10^8$. The minimal and maximal values and the plotted levels are shown in the upper frames. The snapshots at the 1/2 and 3/4 of the time period can be obtained by reversing the colors. The color maps show the temperature field in the characteristic cross-sections. See Supplemental material <Pert_potentials_bif3.avi> for evolution of the perturbation potentials over the time period.

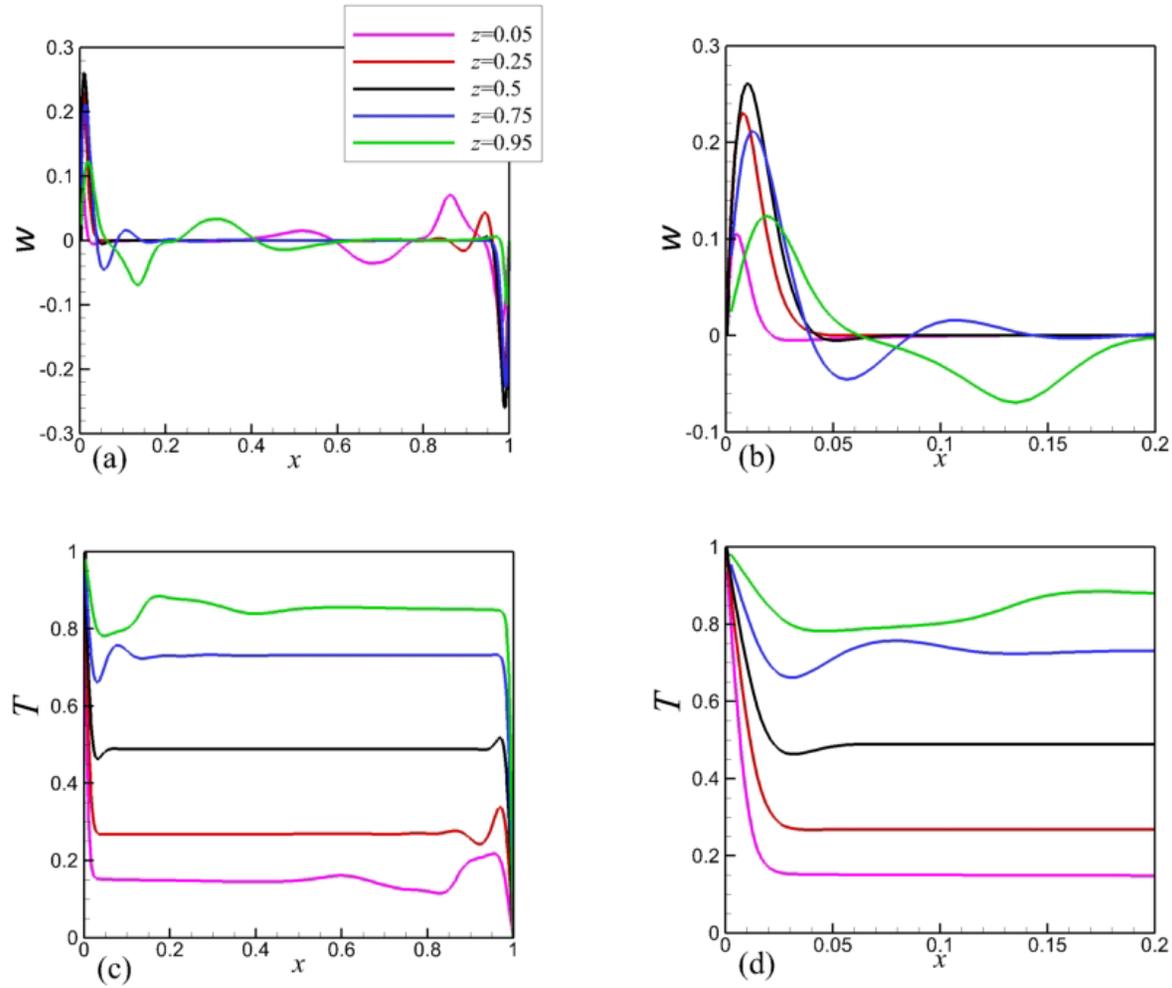

Figure 22. Profiles of the vertical velocity and the temperature in the spanwise midplane $y = 0.5$. In the frames (b) and (d) the profiles in the interval $0 \leq x \leq 0.2$ are zoomed. $Gr = 2.89 \times 10^8$.

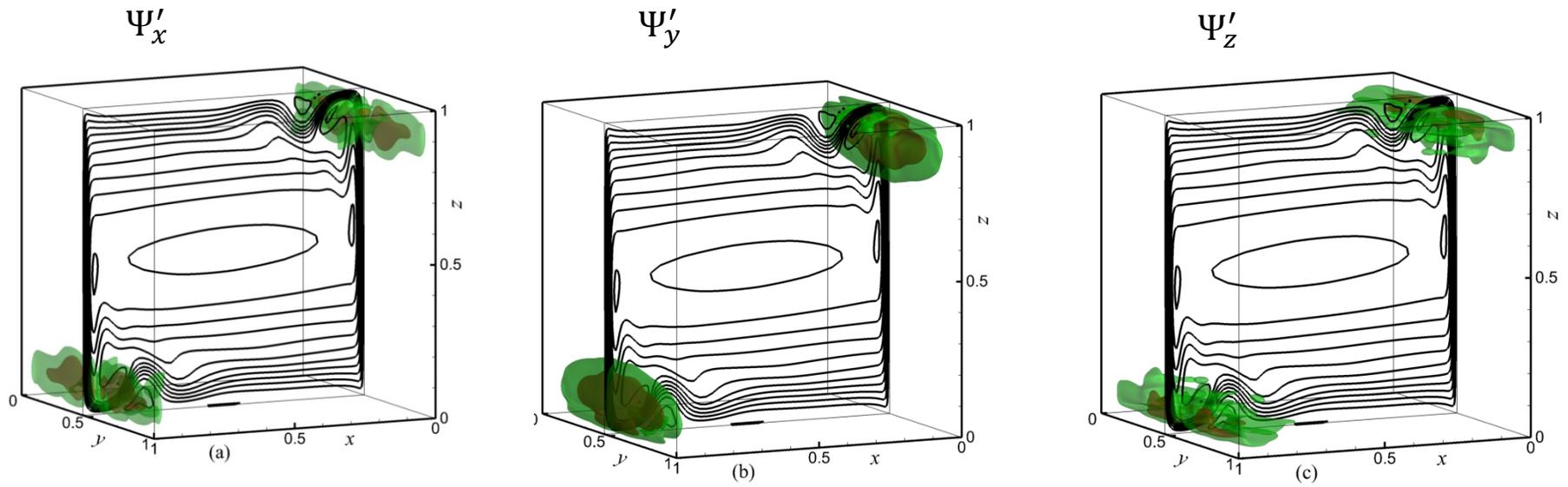

Figure 23. Absolute values of the velocity potentials of the most unstable perturbation of the third bifurcation (isosurfaces) and isolines of the base flow velocity potential $\Psi_y$ plotted in the center plane $y = 0.5$ at $Gr_{cr} \approx 2.96 \times 10^8$. The isosurfaces are plotted for the levels 0.1 and 0.25. The maximal values of $|\Psi'_x|, |\Psi'_y|$ and $|\Psi'_z|$ are 0.640, 0.747, and 0.468, respectively.

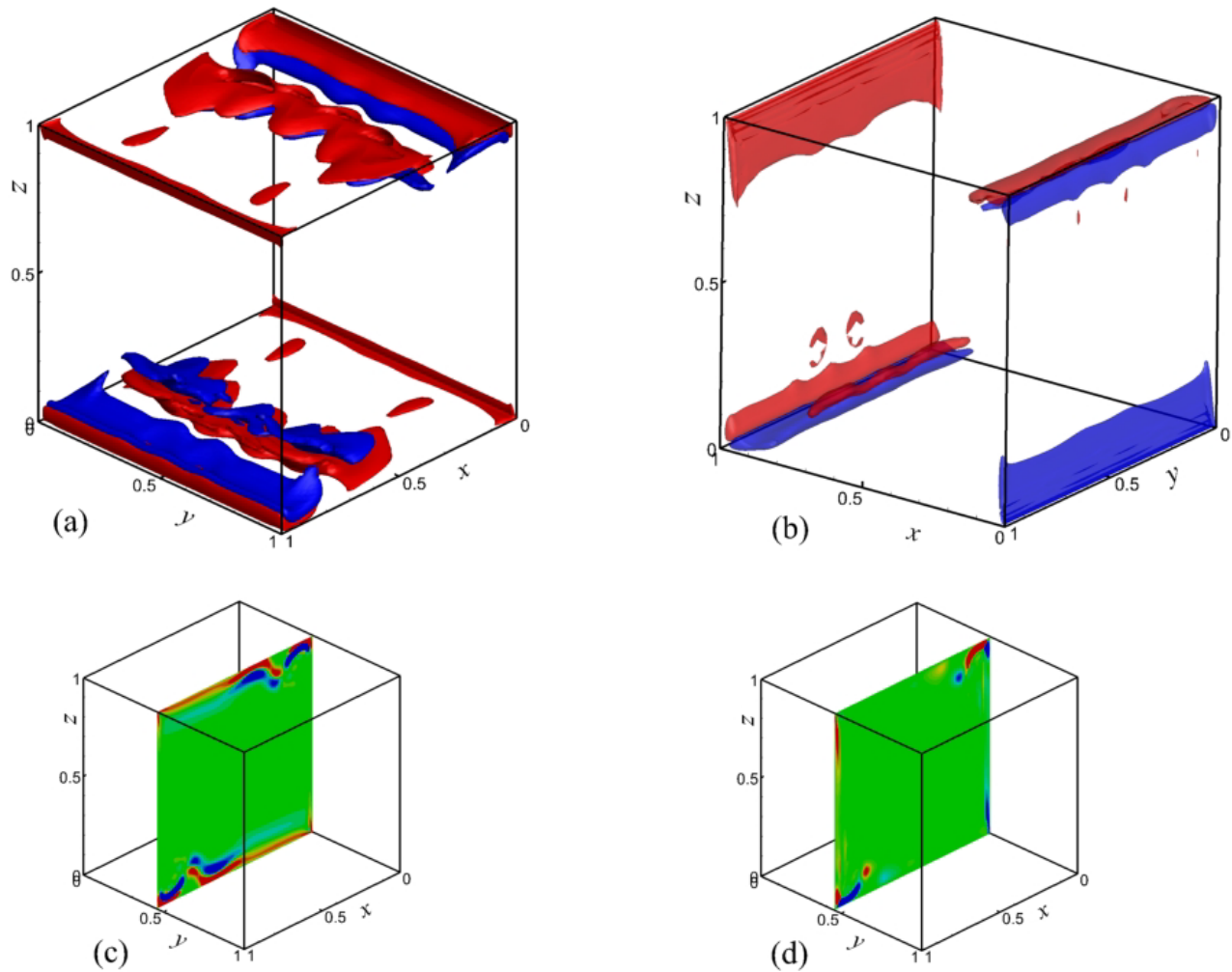

Figure 24. Isolines of Rayleigh (a and c) and Bayly (b and d) criteria calculated for the base flow at $Gr_{cr} \approx 2.96 \times 10^8$. The level values in the frames (a) and (b) are: $\pm 0.03$ for the Rayleigh and $\pm 20$ for the Bayly criteria. Positive and negative values are shown by the red and blue color, respectively.

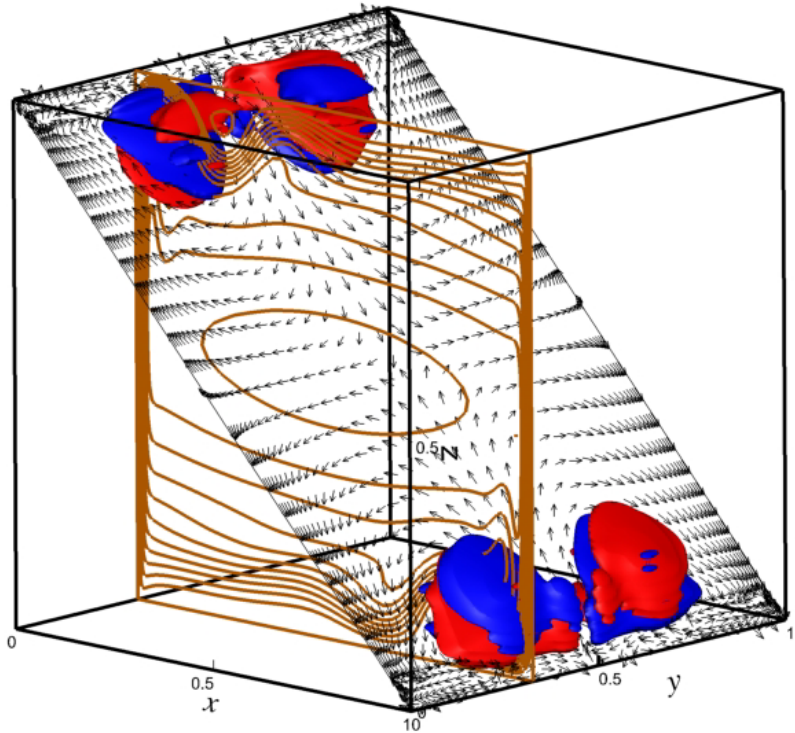

Figure 25. Snapshot of the potential $\Psi'_z(x', y, z')$ obtained after rotation of coordinates by Eq. (10) shown by colors and black isolines for the most unstable mode at $Gr = 2.96 \times 10^8$. The snapshots are shown at the beginning and a half of the oscillations period. The brown isolines illustrate the main convective circulation in the midplane $y = 0.3$ defined by the vector potential of the base flow $\Psi_y$.